\documentclass[iop,appendixfloats, numberedappendix]{emulateapj}

\usepackage{natbib,graphicx, morefloats}
\usepackage{epstopdf}
\usepackage{hyperref}
\shorttitle{X-shooter Spectroscopy of Weak Emission Line Quasars}
\shortauthors{Plotkin et al.}

\begin{document} 
\title{Detection of  Rest-frame Optical Lines from X-shooter Spectroscopy of Weak Emission Line Quasars\footnotemark[$\star$]} \footnotetext[$\star$]{Based on observations made with ESO Telescopes at the La Silla Paranal Observatory under program IDs 088.B-0355 and 090.B-0438}

\author{Richard~M.~Plotkin,\altaffilmark{1}
Ohad~Shemmer,\altaffilmark{2}
Benny~Trakhtenbrot,\altaffilmark{3,13}
Scott~F.~Anderson,\altaffilmark{4}
W.~N.~Brandt,\altaffilmark{5,6,7}
Xiaohui~Fan,\altaffilmark{8}
Elena~Gallo,\altaffilmark{1}
Paulina~Lira,\altaffilmark{9}
Bin~Luo,\altaffilmark{5,6}
Gordon~T.~Richards,\altaffilmark{10}
Donald~P.~Schneider,\altaffilmark{5,6}
Michael~A.~Strauss,\altaffilmark{11}
and Jianfeng~Wu\altaffilmark{12}
}

\altaffiltext{1} {Department of Astronomy, University of Michigan, 1085 South University Avenue, Ann Arbor, MI 48109, USA; \url{rplotkin@umich.edu}}
\altaffiltext{2} {Department of Physics, University of North Texas, Denton, TX 76203, USA; ohad@unt.edu}
\altaffiltext{3} {Institute for Astronomy, Department of Physics, ETH Zurich, Wolfgang-Pauli-Strasse 27, CH-8093 Zurich, Switzerland}
\altaffiltext{4}{Department of Astronomy, University of Washington, Box 351580, Seattle, WA 98195, USA}
\altaffiltext{5}{Department of Astronomy \& Astrophysics, 525 Davey Lab, The Pennsylvania State University, University Park, PA 16802, USA}
\altaffiltext{6}{Institute for Gravitation and the Cosmos, The Pennsylvania State University, University Park, PA 16802, USA}
\altaffiltext{7}{Department of Physics, 104 Davey Lab, The Pennsylvania State University, University Park, PA 16802, USA}
\altaffiltext{8}{Steward Observatory, University of Arizona, 933 North Cherry Avenue, Tucson, AZ 85721, USA}
\altaffiltext{9}{Departamento de Astronomia, Universidad de Chile, Camino del Observatorio 1515, Santiago, Chile}
\altaffiltext{10}{Department of Physics, Drexel University, 3141 Chestnut Street,ÊPhiladelphia, PA 19104, USA}
\altaffiltext{11}{Department of Astrophysical Sciences, Princeton University, Princeton, NJ 08544, USA}
\altaffiltext{12}{Harvard-Smithsonian Center for Astrophysics, MS 6, 60 Garden Street Cambridge, MA 02138, USA}
\altaffiltext{13}{Zwicky Postdoctoral Fellow}


\newcommand{\ha}{H$\alpha$}
\newcommand{\hb}{H$\beta$}
\newcommand{\mgii}{\ion{Mg}{2}}
\newcommand{\ciii}{\ion{C}{3}]}
\newcommand{\civ}{\ion{C}{4}}
\newcommand{\siiv}{\ion{Si}{4}}
\newcommand{\lya}{Ly$\alpha$}
\newcommand{\lyanv}{Ly$\alpha$+\ion{N}{5}}
\newcommand{\feii}{\ion{Fe}{2}}
\newcommand{\feiii}{\ion{Fe}{3}}
\newcommand{\oiii}{[\ion{O}{3}]}
\newcommand{\heii}{\ion{He}{2}}
\newcommand{\uvheii}{\ion{He}{2}~$\lambda$1640}
\newcommand{\optheii}{\ion{He}{2}~$\lambda$4686}
\newcommand{\rewuvheii}{\rew[\heii]$_{\rm uv}$}
\newcommand{\rewoptheii}{\rew[\heii]$_{\rm opt}$}
\newcommand{\rfeiiopt}{R$_{\rm opt, FeII}$}   
\newcommand{\rfeiiuv}{R$_{\rm uv, FeII}$}   
\newcommand{\rciv}{$R_{\rm CIV}$}    

\newcommand{\ledd}{$L_{\rm Edd}$}
\newcommand{\lledd}{$L_{\rm bol}/L_{\rm Edd}$}
\newcommand{\mbh}{$M_{\rm BH}$}
\newcommand{\msun}{$M_\odot$}
\newcommand{\et}{et al.\ }
\newcommand{\kms}{km~s$^{-1}$}
\newcommand{\ergs}{erg~s$^{-1}$}
\newcommand{\rew}{W$_{r}$}
\newcommand{\smbh}{SMBH}

\newcommand{\srcone}{SDSS~J0836}
\newcommand{\srctwo}{SDSS~J0945}
\newcommand{\srcthree}{SDSS~J1321}
\newcommand{\srcfour}{SDSS~J1332}
\newcommand{\srcfive}{SDSS~J1411}
\newcommand{\srcsix}{SDSS~J1417}
\newcommand{\srcseven}{SDSS~J1447}

\newcommand{\srcshonefull}{SDSS J114153.34+021924.3}
\newcommand{\srcshtwofull}{SDSS J123743.08+630144.9}
\newcommand{\srcshone}{SDSS J1141}
\newcommand{\srcshtwo}{SDSS J1237}

\newcommand{\srcwufull}{SDSS J152156.48+520238.5}
\newcommand{\srcwu}{SDSS J1521}

\newcommand{\srcpg}{PG~1407+265}
\newcommand{\srcphl}{PHL~1811}
\newcommand{\srche}{HE 0141$-$3932}

\newcommand{\nfit}{six}
\newcommand{\nfitdig}{6} 
\newcommand{\nsoft}{five}  
\newcommand{\nsmallqso}{104}

\defcitealias{plotkin10}{P10}
\defcitealias{diamond-stanic09}{DS09}

\newcommand{\nar}{New Astronomy Review}

\begin{abstract}
Over the past 15 years, examples of exotic radio-quiet quasars with intrinsically weak or absent broad emission line regions (BELRs)  have emerged from large-scale spectroscopic sky surveys.   Here, we present spectroscopy of seven  such weak emission line quasars (WLQs)  at moderate redshifts ($z=1.4-1.7$) using the X-shooter spectrograph, which provides simultaneous optical and near-infrared spectroscopy covering the rest-frame ultraviolet through optical.   These new observations effectively double the number of WLQs with spectroscopy in the optical rest-frame, and they allow us to compare the strengths of (weak) high-ionization emission lines  (e.g., \civ) to low-ionization lines (e.g., \mgii, \hb, \ha) in individual objects.   We detect broad \hb\ and \ha\ emission in all objects, and these lines are generally toward the weaker end of the distribution expected for typical quasars (e.g., \hb\ has rest-frame equivalent widths ranging from 15--40~\AA.).   However, these low-ionization lines are not exceptionally weak, as is the case for high-ionization lines in WLQs.   The X-shooter spectra also display relatively strong optical \feii\ emission, \hb\ FWHM $\lesssim 4000$~\kms,   and significant \civ\ blueshifts ($\approx$1000-5500~\kms) relative to the systemic redshift; two spectra also show elevated ultraviolet \feii\ emission, and an outflowing component to their (weak) \mgii\  emission lines.  These  properties suggest that WLQs are exotic versions of ``wind-dominated" quasars.  Their BELRs either have unusual high-ionization components, or their BELRs are in an atypical photoionization state because of an unusually soft  continuum.    
 \end{abstract}

\keywords{accretion, accretion disks --- galaxies: active --- quasars: emission lines}

\section{Introduction}
\label{sec:intro}

One of the  prominent observational signatures of Type 1 quasars  is radiation from the  broad emission line region (BELR).  The BELR is composed of high-velocity gas ($\gtrsim$10$^3$~\kms) embedded deep in the gravitational potential well of a quasar's supermassive black hole (SMBH)  ($\lesssim$0.1~pc), which  reprocesses photons from the accretion disk and X-ray corona into Doppler broadened line emission.  The BELR  responds to variations of the quasar's central engine \citep[e.g.,][]{peterson93}, and its properties (e.g., differences in the relative strengths and widths of various emission species, non-virialized motions, etc.)\ are  influenced by fundamental quasar parameters like  luminosity, black hole mass (\mbh) and  Eddington ratio (\lledd)\footnote{$L_{\rm bol}$ is the bolometric luminosity, and \ledd$~= 1.26 \times 10^{38} \left[M_{\rm BH}/M_\odot\right]$~\ergs\ is the Eddington luminosity for ionized hydrogen in a spherical geometry.} 
 \citep[e.g.,][]{baldwin77, boroson92, murray95, murray95a, elvis00, nicastro00, sulentic07, elitzur09, richards11, shen14}.   The BELR is therefore a powerful tool for constraining the energetics of accreting black holes, and for probing the link between SMBHs and their host galaxies.

 An unexpected population of $\sim$10$^2$ quasars were discovered within the  spectra of the Sloan Digital Sky Survey \citep[SDSS;][]{york00} that display exceptionally weak or completely missing broad emission lines in the ultraviolet (UV) rest-frame (\citealt{fan99, anderson01, collinge05}; \citealt{diamond-stanic09}, hereafter \citetalias{diamond-stanic09}; \citealt{plotkin10}, hereafter \citetalias{plotkin10}; \citealt{hryniewicz10}; \citealt{wu11}; \citealt{meusinger14}).  The first systematic search for such weak emission line quasars (WLQs) was performed by \citetalias{diamond-stanic09}, who examined the rest-frame equivalent width (\rew) distribution of the Ly$\alpha~\lambda$1216+\ion{N}{5}~$\lambda$1240 blend for SDSS quasars at $z>3$.  They identified $\sim$70 WLQs, defined  by an excess in the number of high-redshift SDSS quasars with  \rew[\lyanv]$<$15~\AA, which corresponds to the  $>$3$\sigma$ weak tail of the \rew[\lyanv] distribution.  A comparable number of lower-redshift ($z<3$) WLQs  have subsequently been identified within the SDSS as well, with the majority being discovered as a byproduct of searches for BL~Lac objects (\citealt{collinge05}; \citetalias{plotkin10}), as well as through machine learning data mining  techniques \citep{meusinger14} and  serendipitous discoveries  \citep{hryniewicz10, nikoajuk12}.   Most of the $z<3$ WLQs are at too low-redshift for the \lyanv\ blend to be covered by the SDSS spectrograph, and they are generally selected by weak \mgii~$\lambda$2800, \ciii~$\lambda$1909, and/or \civ~$\lambda$1549 (depending on the redshift; see above references for details).  
 
  The puzzle behind WLQs is that they appear to be typical quasars in almost every aspect except for their unusually weak broad emission lines.   The only  property that may hint at a difference from the parent quasar population is that about half of all WLQs are X-ray weaker than expected given their rest-frame UV luminosities \citep{shemmer09, wu11, wu12, luo15}.  Otherwise,  WLQ  optical colors and  variability amplitudes/timescales are similar to other Type 1 SDSS quasars, showing little to no evidence of significant absorption or obscuration by the dusty torus (\citealt{fan99}; \citetalias{diamond-stanic09}).   Dilution of emission lines by a beamed relativistic jet  (as is partly responsible for the weak BELR emission from BL~Lac objects; \citealt{blandford78}) is also extremely unlikely: WLQs are not radio-loud\footnote{Radio loudness is typically determined by the parameter $R=f_{\rm 5~GHz}/f_{\mathrm 4400~\AA}$, where $f_{\rm 5~GHz}$ and $f_{\mathrm 4400~\AA}$ are radio and optical flux densities at 5~GHz and 4400~\AA, respectively.  Radio-loud quasars are usually defined by $R>10-100$ \citep[e.g.,][]{kellermann89}.} 
 (\citealt{collinge05}; \citetalias{diamond-stanic09}; \citetalias{plotkin10}), their optical emission is not polarized (\citealt{fan99, smith07}; \citetalias{diamond-stanic09}; \citealt{heidt11}),  their X-ray to optical luminosity ratios are smaller than for BL Lac objects (\citealt{shemmer09}, \citetalias{plotkin10}, \citealt{plotkin10a, wu11, wu12}) , and their infrared (IR) colors  are also different from BL Lac objects (\citealt{plotkin12}).    Further arguing against a  beamed synchrotron continuum, the UV through IR spectral energy distributions (SEDs) of WLQs are  similar to other SDSS quasars, showing ``big blue bump'' emission in the UV and reprocessed dust radiation in the IR \citep{lane11, wu12}.    WLQs are also far too luminous for their weak or absent BELRs to be explained by a radiatively inefficient accretion flow due to low Eddington ratios, as has been proposed for so-called ``optically dull'' active galactic nuclei (AGN; e.g., \citealt{nicastro03, hawkins04, tran11, trump11}). Finally, there is no evidence that effects from gravitational lensing are important for the weak BELR emission from WLQs (\citealt{shemmer06}; \citetalias{diamond-stanic09}). 
 
Considering the above, WLQs represent one extreme of quasar physics that is  driven predominantly by intrinsic properties of the quasar, and not only by orientation.  Several models have been proposed to explain this extreme population.    These models  fall into two broad categories: (i) WLQ BELRs are unusually gas deficient (we  refer to this as the \textit{anemic} BELR idea);  and (ii) WLQ BELRs  are in an unusual ionization state, most likely caused by a softer than normal ionizing continuum (which we refer to as the \textit{soft continuum} idea).  For the anemic category, WLQs could simply have low gas content and/or covering factors \citep[e.g.,][]{shemmer10, nikoajuk12}.   In this case, we may expect all broad lines to have low equivalent widths and luminosities.   Another potential explanation within the anemic category is that WLQs are just beginning a quasar phase, and their BELRs are bare because a disk wind has not yet had sufficient time to lift material from the accretion disk into the BELR \citep{hryniewicz10}.   Assuming disk wind velocities of $\approx$10$^2$~\kms,  \citet{hryniewicz10} propose that BELR formation should take  $\sim$10$^3$~yr and WLQs  should therefore be rare.  During the WLQ phase, low-ionization emission lines that form close to the accretion disk (e.g., \hb, \mgii) may appear relatively normal, while higher-ionization lines with ``wind'' components (e.g., \civ)  appear weakest because they originate in a region higher above the disk that has not yet fully formed.   

The soft continuum category can be subdivided into physical mechanisms that produce intrinsically soft continua, and those that modify the  continuum  prior to illuminating the BELR (e.g., through absorption).   One way to produce an intrinsically soft continuum is through a cold accretion disk.   \citet{laor11} show that, since the temperature of a standard thin accretion disk scales as \mbh$^{-0.25}$, extremely massive SMBHs (\mbh~$>3.6\times10^9$~\msun\ for a non-spinning black hole) would produce an accretion disk too cold to photoionize the BELR.  With very massive SMBHs, one expects all emission lines to be weak and broad.  Another way to form an intrinsically soft continuum is through Super-Eddington accretion, as has been proposed for the nearby ($z=0.19$) narrow-line  Seyfert 1 galaxy \srcphl, which has exceptionally weak high-ionization lines (i.e., \rew[\civ]= 6.6~\AA) and is intrinsically X-ray weak \citep{leighly07a, leighly07}.  Super-Eddington accretion could produce softer ionizing continua through less efficient X-ray production, perhaps related to a smaller or quenched X-ray corona, or X-ray photons could be advected into the black hole.  Regardless of the exact physical reason, the effect is that super-Eddington accretion could sometimes be associated with a more narrow, UV-peaked continuum that does not emit enough high-energy photons to form high-ionization potential species (like \civ); however, lower-ionization species (e.g., \ha, \hb, \mgii) would still be normal \citep{leighly07a, leighly07}.  

  A final mechanism proposed by \citet{wu11} invokes an ``X-ray shielding'' gas that is located interior to the BELR.  \citet{wu11} examined the X-ray properties of an optically selected sample of 10 $z\sim2.2$  SDSS quasars that have similar rest-frame UV spectral properties as  \srcphl\ \citep[namely, weak and blueshifted \civ, and strong UV Fe emission;][]{leighly07}.  Compared to the parent quasar population, their \srcphl-analogs display lower average X-ray to UV luminosity ratios (by a factor of 13) and  harder average X-ray spectra.  To explain these observations,   \citet{wu11} propose that an unusually large amount of X-ray absorbing gas exposes the BELR to  a softer than typical continuum.   \citet{wu11} then suggest that their \srcphl-analogs could be a subset of WLQs, oriented such that the X-ray continuum is  viewed through a column of this shielding gas; an X-ray normal WLQ would be observed at lower inclinations \citep[also see][for further observational support for this idea]{wu12}. The ratios of the strengths of high- to low-ionization emission lines is expected to be similar in both X-ray normal and weak  objects, regardless of orientation, and we generally might expect high-ionization lines to be weaker than low-ionization lines.

  Differentiating between the above scenarios requires multiwavelength observations to constrain the SED, and also broad spectral coverage in order to compare the relative strengths of high- and low-ionization  emission lines.  The X-ray properties of WLQs are discussed in another publication (\citealt{luo15}; also see \citealt{shemmer09, wu11, wu12}).  Here, we focus on the UV--optical spectral properties of WLQs.  Unfortunately, the existing SDSS spectra of WLQs provides too narrow of a spectral window to place meaningful constraints on the above models, and near-infrared (NIR) spectroscopy  is required to study the rest-frame optical.  We therefore undertook a campaign with the X-shooter spectrograph \citep{vernet11} on the Very Large Telescope (VLT), targeting a subset of seven WLQs at $1.4 \lesssim z \lesssim 1.7$.    Prior to these X-shooter observations, only two other SDSS WLQs (both at $z\sim3.5$; \citealt{shemmer10}), and one \srcphl-analog ($z\sim2.2$; \citealt{wu11}) had   rest-frame optical constraints via NIR spectroscopy.   In Section~\ref{sec:obs} we describe our X-shooter observations and data reduction.  Spectral fits to emission lines are described in Section~\ref{sec:specanal}, and samples of comparison quasars are assembled in Section~\ref{sec:qsos}.  We then present our results in Section~\ref{sec:res}, which are  discussed in Section~\ref{sec:disc}.  Throughout, we adopt $H_0=70$~km~s$^{-1}$~Mpc$^{-1}$, $\Omega_\Lambda=0.7$, $\Omega_M$=0.3.   We define radio-quiet quasars as having a radio-loudness parameter $R< 10$, and error bars are reported at the 68\% confidence level, unless stated otherwise.  

\section{X-shooter Observations}
\label{sec:obs}

\subsection{Sample Selection} 
\label{sec:obs:samp}
 We focus on WLQs at moderate redshift ($z\sim1.5$) here, a redshift where X-shooter covers a large range of emission lines (spanning  \civ\ through \ha).  We select  targets from the lists of radio-quiet, weak-featured quasars  in Table~6 of \citetalias{plotkin10} (86 objects) and Table~5 of \citet[][27 objects]{collinge05}.  These catalogs  include  quasars that pass  \textit{optical} spectroscopic criteria to be classified as BL~Lac objects, namely that all line emission  has \rew$<$5~\AA,  and the 4000~\AA\ break is smaller than 40\% if present \citepalias[see][for details]{plotkin10}.   To select X-shooter targets, we restrict the above lists to sources with declinations $\delta \lesssim 20^\circ$ (to be visible by the VLT),  we require targets to have a reliable redshift  (from weak emission features in their SDSS spectra, typically \mgii, and sometimes \ciii\ and \civ) in order to avoid any contamination from stars, and to ensure that the \hb\ line will fall in a NIR atmospheric transmission window, and we require sufficient signal-to-noise ($S/N$) in $<$1 hour  observing blocks.  These cuts ween our target list to $\sim$15 sources. 
  
 For observations,  we select six  targets ($1.4<z<1.7$) with SDSS spectra that are representative of the larger $\sim$15 object sample (in terms of relative line strengths and continuum shapes).  We also include the $z=1.7$ WLQ   SDSS~J094533.98+100950.1 (hereafter \srctwo), which was discovered serendipitously by \citet{hryniewicz10}.  We include \srctwo\ because it has slightly stronger \mgii\ emission than the other targets, and it therefore probes a  different part of  WLQ parameter space  (also see, e.g., \citealt{laor11, czerny11a} for constraints this particular source has placed on quasar accretion disks).  Observations of these seven WLQs were taken over two observing seasons (program IDs 088.B-0355 and 090.B-0438; PI Plotkin), and a summary of the observations is provided in Table~\ref{tab:obslog}.    Throughout the text, we refer to each source by its SDSS designation truncated to   the first four digits.

\subsection{Observations and Data Reduction}
\label{sec:obs:datared}
 Similar instrument configurations and observing strategies were employed for all seven observations.  X-shooter  splits the incoming light beam into three segments or `arms'  (termed UVB, VIS, and NIR) using two dichroics at 560 and 1024 nm, and the light is  fed into three independently operated spectrographs.    We used a $1.0\times11.0 \arcsec$ slit for the UVB arm and $0.9\times11.0 \arcsec$ slits for the VIS and NIR arms, providing spectral resolutions of $R\sim$4350, 7450, and 5300 in the UVB, VIS, and NIR arms, respectively.  The detectors were  binned by 2 pixels along the dispersion direction for the UVB and VIS arms, and no binning was used for the NIR arm.  We acquired four exposures for each target, nodding the target along the slit in an ABBA pattern (4.5$\arcsec$ separation between positions).  We observed \srctwo\ and \srcfour\ for $\sim$20 min each, and each of the other five targets were observed for $\sim$40 min (see Table~\ref{tab:obslog}).  To improve $S/N$ in the NIR arm, the four observations from 2013 were taken with the K-band blocking filter in place (the blocking filter was not available at the time of the other three observations).  For telluric corrections, an A0V star was observed before or after each WLQ  at a similar airmass, with the same slit and instrument configuration.  A spectrophotometric standard star was observed for flux calibration at the beginning or end of each night using a 5$\arcsec$ slit  (the flux standard was taken on the following night for \srcthree\ and \srcsix).  Each X-shooter arm is reduced individually using the X-shoter pipeline \citep{modigliani10}, following standard procedures that are described in detail in the Appendix. The final flux calibrated spectra (including telluric corrections, and corrections for Galactic extinction) are presented in Figure~\ref{fig:fullspec}.

\begin{figure*}
\centering
\includegraphics[scale=0.82]{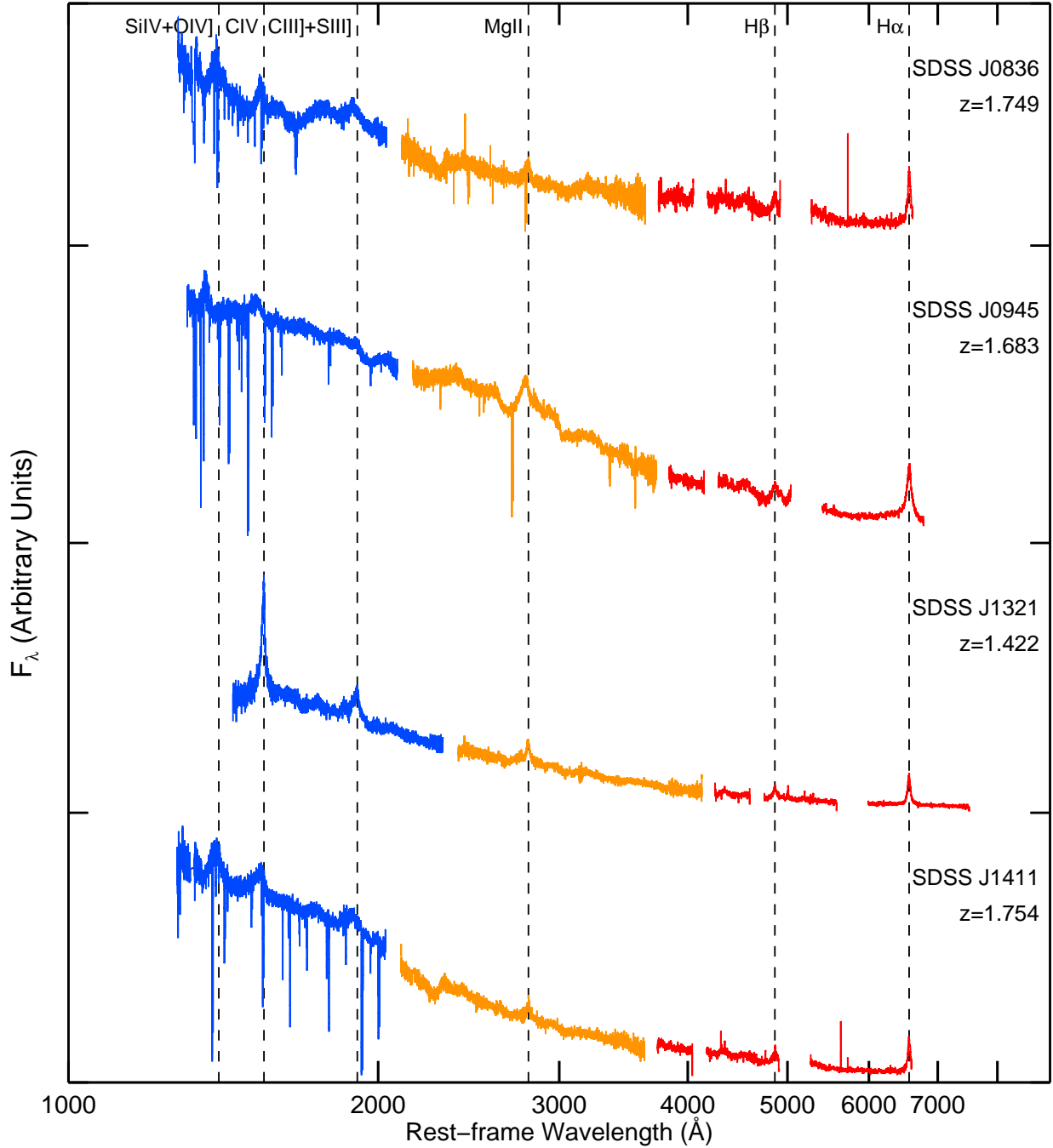}
\caption{Full X-shoter spectra. The UVB, VIS, and NIR arms are shown in blue, orange, and red, respectively.  Resolution elements are  $\sim$2-3~\AA\ in each arm.  Positions of prominent emission lines are indicated.}
\label{fig:fullspec}
\end{figure*}

\begin{figure*}
\centering
\includegraphics[scale=0.82]{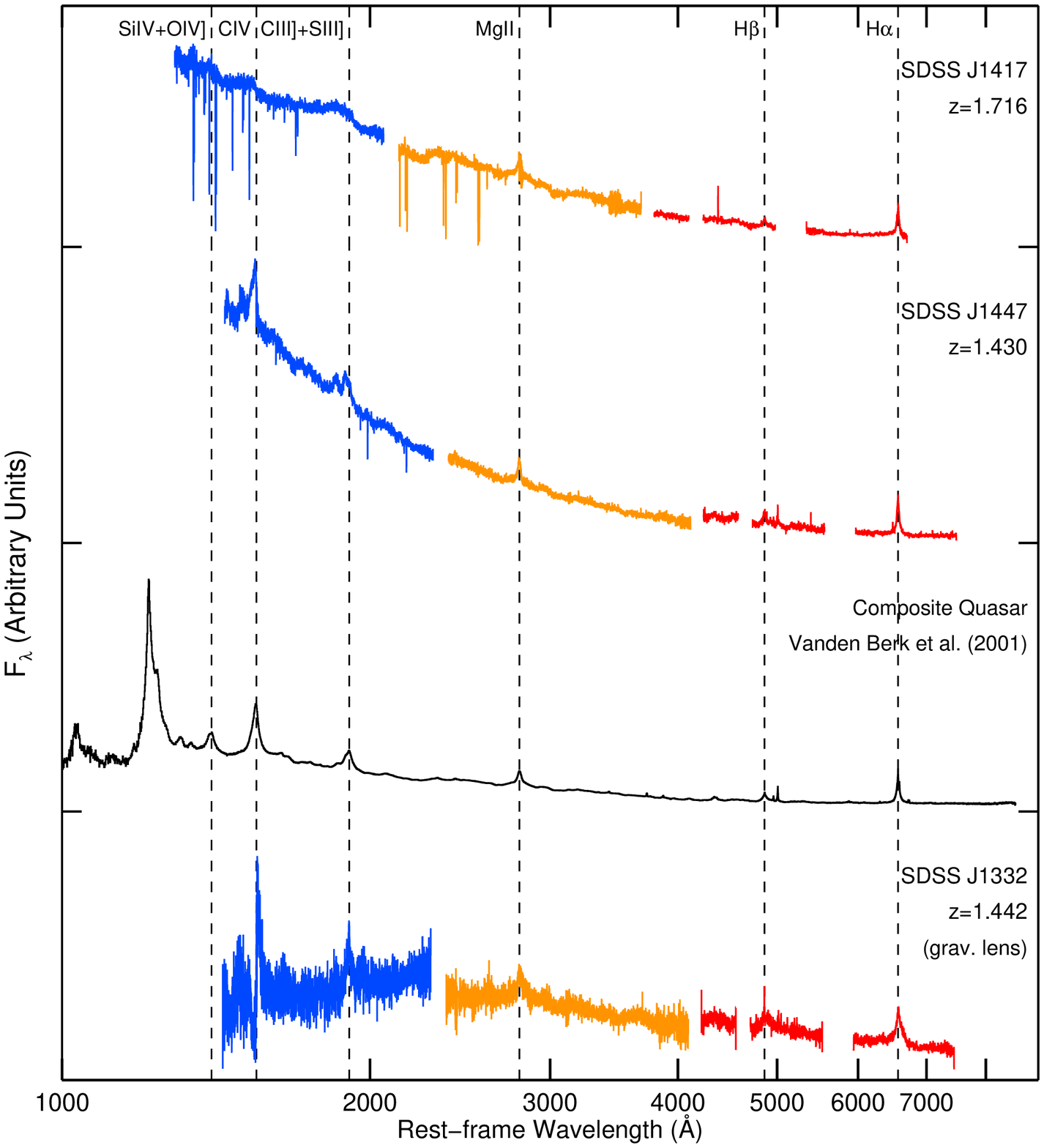}
\addtocounter{figure}{-1}
\caption{(Cont'd)  The bottom two spectra are the composite SDSS quasar spectrum from \citet{vanden-berk01} for comparison, and \srcfour\ (for completeness) which is a gravitationally lensed quasar.}
\end{figure*}

Absolute flux calibration of Echelle spectroscopy is difficult, especially for an instrument like X-shooter with a very broad spectral range \citep[see, e.g.,][for a recent discussion specific to X-shooter]{pita14}.   We crudely estimate the accuracy of our flux calibration by comparing the luminosity at 5100~\AA, $L_{5100}=\left(\nu L_\nu\right)_{5100}$, in our flux calibrated  X-shooter spectra to $L_{5100}$ inferred from SDSS photometry (extrapolating from the psf $i$-band magnitude, and assuming a continuum following $f_\nu \propto \nu^{-0.5}$; e.g., \citealt{vanden-berk01}).   The X-shooter luminosity at 5100~\AA\ is typically 20--50\% lower  than in the SDSS, except for \srcone\ where X-shooter is brighter by $\sim$10\%.    We suspect that variability could contribute up to $\sim$15\% of the difference in flux, based upon a comparison of the $i$-band psf magnitude from the SDSS photometric epoch to the synthetic $i$-band magnitude from the SDSS spectroscopic epoch for each source.  Of course, we do not expect variability to systematically bias the flux calibration in the same direction for all sources.   Regardless, we suspect that our X-shooter flux calibration is systematically biased toward lower values for most sources (although the magnitude of the bias is not well-determined).  This bias could in turn  influence measurements of line luminosities, black hole masses, and Eddington ratio estimates.   We stress, however, that \rew\ and FWHM measurements are not  affected by uncertainties in the flux calibration.

\subsubsection{SDSS~J133222.62+034739.9}
\label{sec:gravlens}
\srcfour\ ($z=1.4$) is a known, doubly-imaged gravitationally lensed quasar \citep{morokuma07}.   Even though there is currently no compelling evidence that lensing is related to the WLQ phenomenon (see, e.g., \citealt{shemmer06}; \citetalias{diamond-stanic09}), \srcfour\ provides an  interesting test case to further investigate any potential relationship between lensing and WLQs.  For example, one may expect extinction from the foreground galaxy, and/or different magnifications of the quasar continuum compared to the BELR due to their different physical sizes.  The SDSS spectrum of \srcfour\ does not appear remarkably different than other SDSS WLQs.  However,  X-shooter provides bluer wavelength coverage than the SDSS, and from X-shooter it is clear that the UV continuum of \srcfour\ is redder than the other  targets, and an absorption trough is seen blueward of \civ.  It is not obvious if the different continuum is predominately intrinsic to the quasar (i.e., associated dust obscuration) or related to the gravitational lens.  Regardless, this source illustrates that lensing unlikely contributes to the unusual spectral features of WLQs, since a quasar known to be lensed displays different properties than the rest of the population.  We include the X-shooter spectrum of \srcfour\ in Figure~\ref{fig:fullspec} and Table~\ref{tab:obslog} for completeness.  However, we exclude \srcfour\ from further analysis, to avoid any possible systematics due to extrinsic effects related to the lens.

\section{Spectral Analysis}
\label{sec:specanal}

\subsection{Systemic Redshifts}
\label{sec:redshift}
Prior to these X-shooter spectra, our targets' redshifts were estimated from rest-frame UV emission lines in their SDSS spectra.  The SDSS-derived redshifts are listed in Table~\ref{tab:obslog}, as taken from \citet{hewett10}.   UV emission lines can be susceptible to blueshifts (relative to their laboratory wavelengths) due to winds or other non-viralized motion.  From the X-shooter NIR spectra, we  derive systemic redshifts ($z_{\rm sys}$) from lower-ionization emission lines, which have profiles that are expected to be dominated be virialized motion.  Systemic redshifts are determined, in order of preference, from the peak of narrow [\ion{O}{3}] (which is firmly detected only for \srcthree) or from the average of the peaks of broad \ha\ and \hb.  We estimate that we can  determine the wavelength of the peak line flux density to within one resolution element ($\sim$3~\AA\ in the NIR), so that typically $\sigma_{z_{\rm sys}}  \lesssim 0.0006$.  The new systemic redshifts are similar to the SDSS redshifts from \citet{hewett10} except for two sources, where the new systemic redshifts are slightly larger  by $\Delta z=0.012$ (\srctwo) and $0.006$ (\srcsix).  The  X-shooter derived systemic redshifts are  listed in Table~\ref{tab:obslog} and are adopted as the redshift for each source throughout the rest of the text.

\subsection{Continuum Shape}
\label{sec:pl}
We fit a power law ($f_\lambda \sim \lambda^{\alpha_\lambda}$) to each X-shooter spectrum in order to constrain  the shape of the UV continuum.  The power law is fit between the following (rest-frame) wavelength regions, 1680-1710, 1975-2050, 2150-2250, and 4010-4050~\AA, by using a $\chi^2$ minimization routine (the fit is performed to all measured flux densities within each spectral window, weighted by the uncertainty on each measurement). The above spectral windows are relatively free from contamination from emission lines (including blended iron emission), they are common between all seven X-shooter targets (given their redshifts), and they provide a wide dynamic range.   The best-fit spectral indices are given in Table~\ref{tab:obslog} and, excluding \srcthree, they  are typical of other SDSS quasars (which  have  $\alpha_{\lambda} \approx -1.56$; \citealt{vanden-berk01}). 

\subsection{Line Fitting}
\label{sec:fits}
\begin{figure*}
\centering
\includegraphics[scale=0.82]{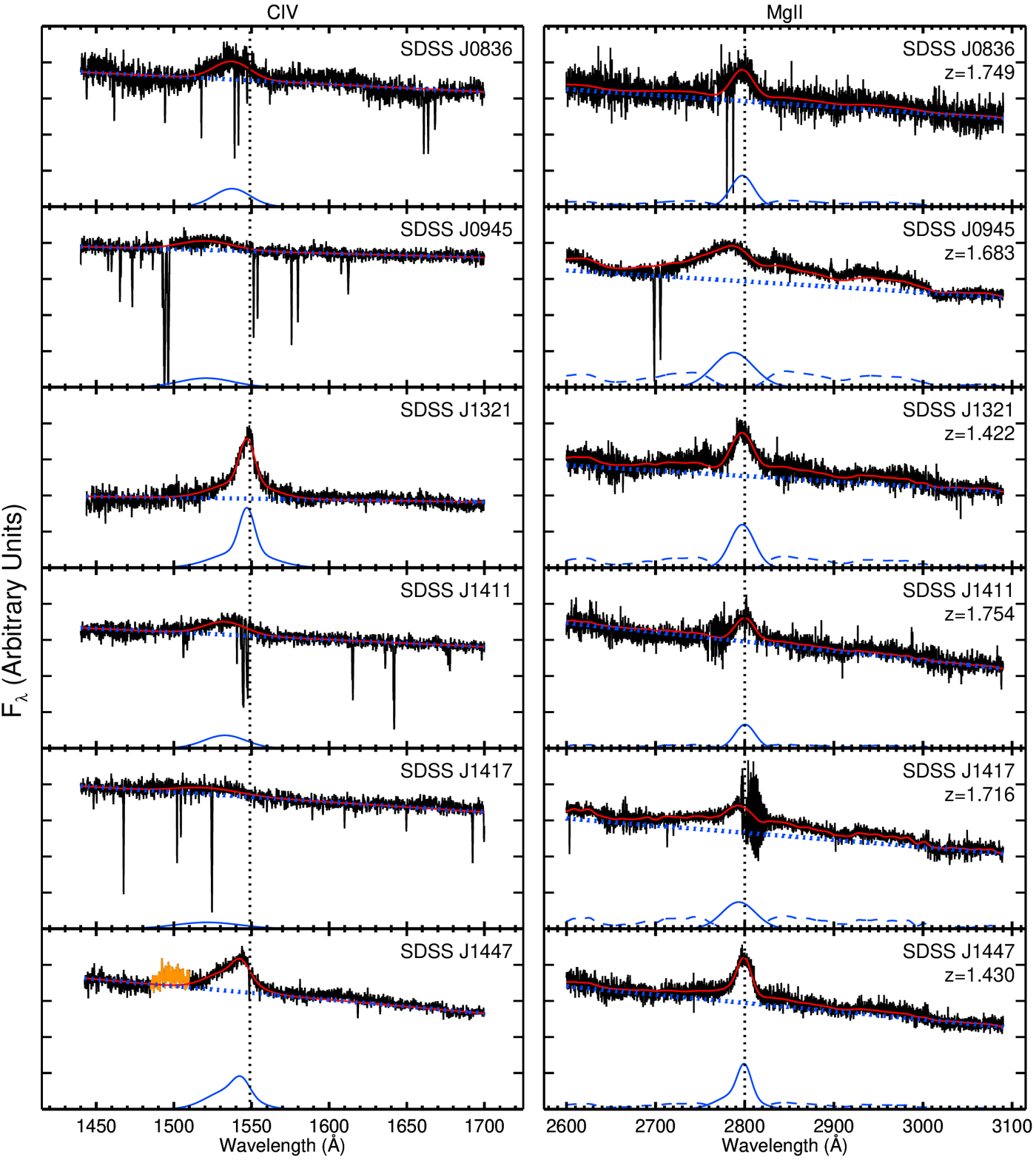}
\caption{Expanded views of the \civ\ (left columns) and \mgii\ (right columns) complexes and their spectral fits.  The same source is shown in each row.  Included in each panel are the best-fit emission line profiles (solid blue line), the best-fit linear continuum (dotted blue line), and the Iron continuum (dashed blue line; when required by the fit). The vertical dotted line indicates the rest-frame wavelengths of \civ\ and \mgii, to help visualize when those lines are blueshifted relative to the systemic redshift of each quasar.  Narrow absorption line systems were masked out during the spectral fitting.  The orange segment for \civ\ of \srcseven\ is a calibration artifact and is not included in the spectral fit (see Section~\ref{sec:fits:notes}).  }
\label{fig:linefitstwo}
\end{figure*}

\begin{figure*}
\centering
\includegraphics[scale=0.82]{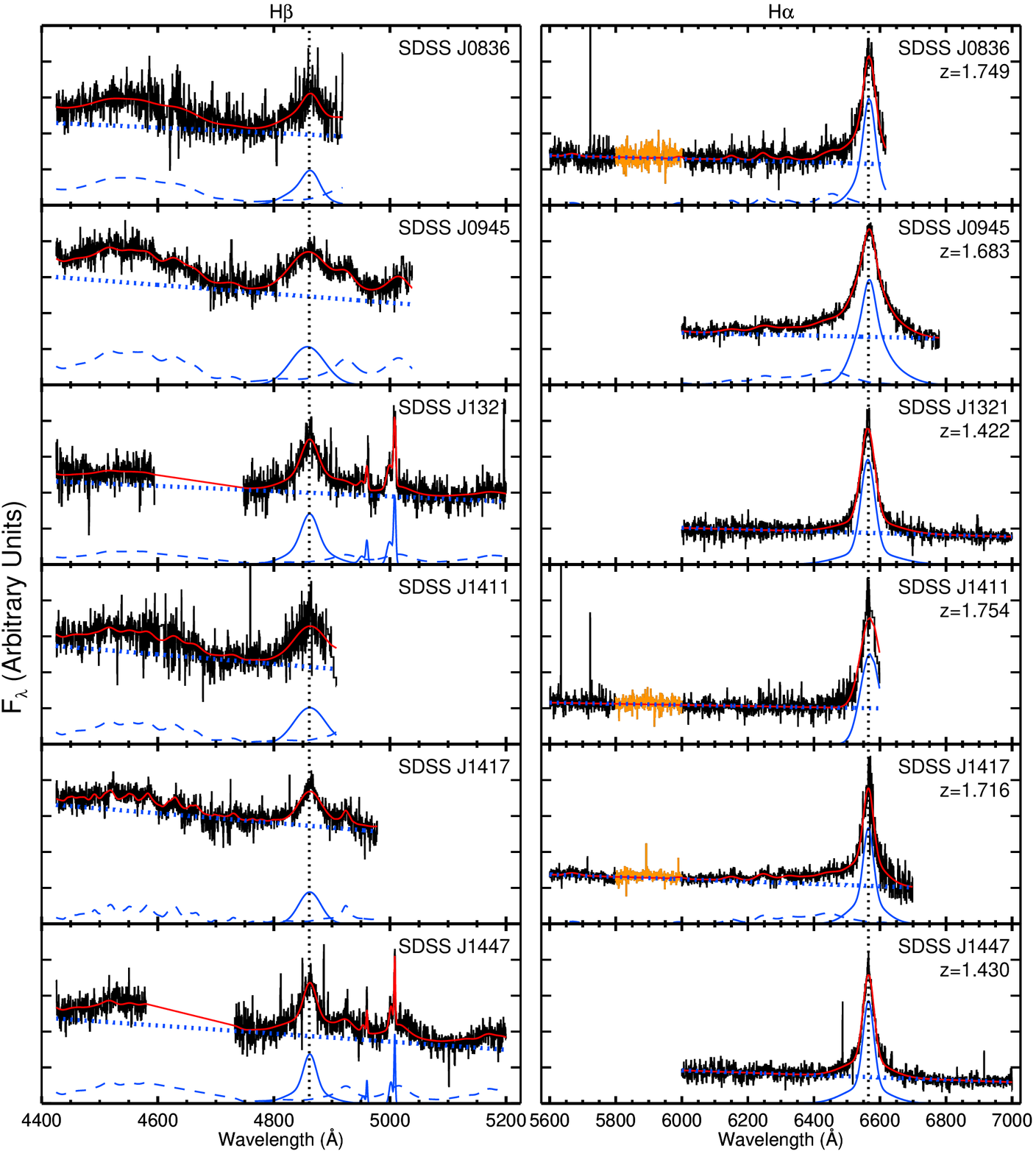}
\caption{Expanded views of the \hb\ (left column) and \ha\ (right column) complexes and their spectral fits.   The format is the same as in Figure~\ref{fig:linefitstwo}.   Narrow Gaussians are included in the spectral fits to model \oiii\ for \srcthree\ and \srcseven. The orange segments denote wavelengths omitted from the \ha\ spectral fits for \srcone, \srcfive, and \srcsix\ in order to avoid potential (very weak) \ion{He}{1} $\lambda$5877 emission  (see Section~\ref{sec:fits:notes}). }
\label{fig:linefits}
\end{figure*}

For each of the \nfit\ WLQs,  we perform spectral fits to the \ha, \hb, \mgii, and \civ\  complexes,  fitting each complex separately by following a procedure similar to \citet{trakhtenbrot12}. These are among the strongest emission lines in quasars, and the ones that  can be measured most accurately given the $S/N$ of our data.   Our spectral model includes a local linear continuum,  (up to three) Gaussians to model each  emission line, and templates for broadened \feii\ and \feiii\ emission  (based on \citealt{boroson92} in the optical and \citealt{vestergaard01} in the UV; see Appendex C of \citealt{trakhtenbrot12} for details).    Details on the spectral fitting (including how we associate uncertainties to best-fit parameters) are provided in the Appendix.    For the \hb\ complex of \srcthree\ and \srcseven, we include narrow Gaussians to model the \oiii\ $\lambda$4959, 5007 emission lines.  These are the only two targets with sufficient spectral coverage to include \oiii\ profiles in the spectral modeling (due to their relatively lower redshifts).  We also attempted to include other narrow emission line species that fall within each complex, but doing so did not improve the quality of any other fits (see Appendix).  The best-fit models to each complex are shown in Figures~\ref{fig:linefitstwo} and \ref{fig:linefits}, and the corresponding spectral properties (i.e., \rew, FWHM, and line luminosities) are presented in Tables~\ref{tab:rew}--\ref{tab:linelum}.

\section{Samples of Comparison Quasars}
\label{sec:qsos}

\subsection{Additional WLQs with NIR Spectroscopy}
\label{sec:res:otherwlqs}
Only three other SDSS WLQs have NIR spectroscopy, all of which are at higher redshift than our X-shooter targets.  These  sources include  \srcshonefull\ (hereafter \srcshone; $z=3.55$) and \srcshtwofull\ (hereafter \srcshtwo; $z=3.49$) presented by \citet{shemmer10}, and \srcwufull\ (hereafter \srcwu; $z=2.238$) presented by \citet{wu11}.   We note that \citet{wu11} identified \srcwu\ as a \srcphl-analog, and it is very X-ray weak (by a factor of 34.5) and extremely optically luminous ($M_i=-30.19$~mag; \citealt{just07}).

 We also consider the  quasar \srcpg\ ($z=0.94$), which we  consider to be the ``prototype'' WLQ because it displays similar properties to the WLQ population later identified by the SDSS.  \srcpg\ was identified by \citet{mcdowell95} as having an unusual BELR, with extremely weak high-ionization emission lines but otherwise normal quasar properties. We also consider  \srcphl\ ($z=0.19$), another AGN with unusual BELR properties similar to WLQs (see \citealt{leighly07a} for a broad spectrum covering \lya\ through \ha).   

There is  not uniform data coverage among the five quasars listed above.  We therefore consider only  their \civ\ and \hb\ equivalent widths, which are included in Table~\ref{tab:rew} for reference.  These five sources generally show relatively normal low-ionization emission lines, weak high-ionization lines, and often  blueshifted \civ\ emission.

\subsection{Samples of Parent Quasars}
\label{sec:res:compqso}
We create samples of other quasars, in order to compare  our WLQs to the parent population.   We start with the SDSS Data Release 7 (DR7) quasar catalog \citep{schneider10} and the emission line measurements provided by \citet{shen11}.  We restrict the SDSS quasar sample to only include quasars targeted for SDSS spectroscopy using the \citet{richards02} algorithm, which is flux limited to $i=19.1$~mag, and we also exclude all objects identified as weak-featured quasars in \citet{collinge05} and \citetalias{plotkin10}.  We only include quasars with absolute $i$-band magnitudes $M_i>-28$~mag (K-corrected to $z=2$),  to compare to quasars that are of similar luminosities as our X-shooter targets.  We also require quasars to not be flagged as broad absorption line quasars (BALs) by \citet{shen11}, and to not be radio-loud (as determined by \citealt{shen11}).  

\subsubsection{The ``Intermediate-redshift'' Sample}
\label{sec:res:dr7iz}
For comparisons to rest-frame UV emission lines, we consider the subset of the above quasars at $1.5<z<2.0$.  SDSS spectra of quasars in this redshift range include both \civ\ and \mgii, allowing a comparison of both emission species within individual objects.  We further remove quasar spectra with large fractions of bad pixels within each emission complex, requiring $>$250 and $>$300 usable pixels for \civ\ and \mgii, respectively.   The final sample includes 10956 quasars with $-28.00<M_i<-25.87$~mag (median $M_i=-26.6$~mag).  We refer to these quasars as the ``SDSS intermediate-redshift'' (SDSS-IZ) subset.  

\subsubsection{The ``Low-redshift'' Sample}
\label{sec:res:dr7lz}
To compare to the \hb\ spectral regime, we assemble a subset of comparison quasars at $0.4<z<0.85$.  This redshift range covers both the \hb\ and the \mgii\ complexes.  The overlap with \mgii\ is useful for comparing line ratios within individual objects, and also to investigate potential biases introduced from comparing lower-redshift (and typically less luminous) quasars to our higher-redshift X-shooter targets (i.e., by comparing \mgii\ properties between this subset and the SDSS-IZ subset).  We require the \mgii\ and \hb\ spectral complexes have $>$300 and $>$150 good pixels, leaving 8151 quasars with   $-27.76<M_i< -22.97$~mag (median $M_i=-24.49$~mag).  We refer to these quasars as the ``SDSS low-redshift'' (SDSS-LZ) subset.

\subsubsection{The ``Small'' Sample}
There are only $\sim$10$^2$ quasars in the literature with high-quality spectra covering both the rest-frame UV and optical.  In turn, the above SDSS-IZ and SDSS-LZ subsets represent our best means to \textit{statistically} compare our targets to the parent quasar population.   However, the drawback is that we cannot compare among individual objects the full range of emission species covered by the X-shooter spectra.  To address this shortcoming, we also compile a sample of $\sim$10$^2$ quasars from the literature that  have high-quality spectral coverage of both \civ\ and \hb.    

We start with the catalog of line measurements for 85 quasars presented by \citet{tang12}.  Their sample is based on \citet{shang11}, and includes UV-selected Palomar-Green (PG) quasars at $z<0.4$, UV-bright quasars at  $z<0.5$ with Hubble Space Telescope (HST) UV spectroscopy, and a subset of radio-loud quasars extending out to $z\sim1.5$.  We include 27 of their $z<0.5$ quasars that are not known to be radio-loud or BAL quasars.  We also include quasars from the study of \citet{baskin04}, who investigated the \civ\ and \hb\ properties of 81 PG quasars at $z<0.5$.  We include 41 of those quasars, after removing known radio-loud quasars, BAL quasars, and sources that also appear in \citet{tang12}.  For these 41 quasars, we adopt the \civ\ measurements presented by \citet{baskin04}, and  \hb\ measured on the original optical data by \citet{trakhtenbrot12}.  Finally, we also include 36 quasars from \citet{shemmer04} and \citet{netzer07a}, who obtained rest-frame optical spectroscopy of  $z\sim2.4$ and 3.3 quasars.  We remove known radio-loud quasars and BAL quasars from their samples, and we adopt the \hb\ measurements performed on this subset by \citet{trakhtenbrot12}, and \civ\ measurements by \citet{shemmer15}.   In total, we assemble \nsmallqso\ quasars from the literature with \hb\ and \civ\ coverage, which we refer to as the ``small'' quasar sample.

\begin{figure*}
\includegraphics[scale=0.7]{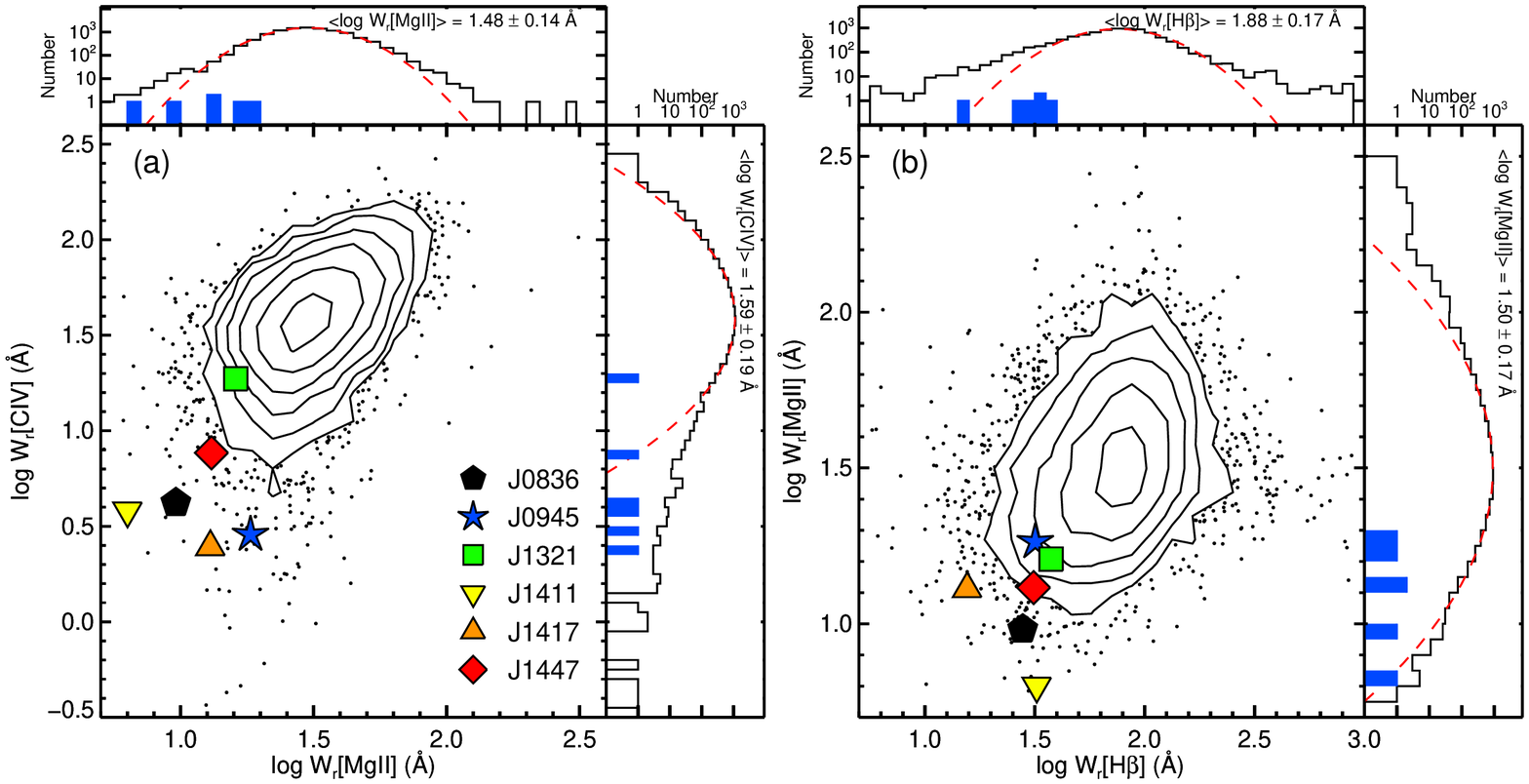}
\caption{Logarithms of the rest-frame equivalent widths (\rew) of our six X-shooter targets (filled symbols) compared to SDSS quasars (contours and black circles; line measurements for the comparison quasars are taken from \citealt{shen11}).  (a) \civ\ vs.\ \mgii; comparison quasars are the SDSS-IZ sample.  (b) \mgii\ vs. \hb; comparison quasars are  the SDSS-LZ sample.  The lowest contour level in each panel denotes 10 quasars per bin ($25 \times 25$ bins along each axis), and increasing contours are logarithmically spaced by 0.35~dex; black points show individual quasars that fall below the lowest contour level.  The subpanels on the top and right sides of panels (a) and (b) display histograms of the comparison quasars (black line), with the best-fit log-normal distribution plotted as a red dashed line.  X-shooter WLQs are shown as filled blue histograms.  Although all WLQ  emission lines are weaker than typical quasars, only \civ\ is substantially weaker at the $>$3$\sigma$ level (except for \srcthree). }
\label{fig:ewVew}
\end{figure*}

\section{Results}
\label{sec:res}
Here, we present the X-shooter spectral properties of our  \nfit\ WLQs, with  physical interpretations discussed in Section~\ref{sec:disc}.   The NIR arm provides our first view of the rest-frame optical properties of these WLQs, and we  focus on \hb\ (not \ha) because we do not have a statistical sample of comparison quasars covering both high-ionization emission lines and \ha.  Besides the extended coverage from the NIR arm, the UVB and VIS arms represent improvements over the extant SDSS spectra because of their higher spectral resolution, and because the UVB arm extends to shorter wavelengths than the SDSS.  

All broad emission lines from our \nfit\ WLQs have \rew\ values that fall toward the weaker end of the distributions for the SDSS comparison quasars  (see Figure~\ref{fig:ewVew}).   The X-shooter targets have relatively weak \mgii\ and \hb, but those \rew\ measures still fall within a range  that is commonly displayed by the parent SDSS quasar population.  On the other hand, \civ\ approaches a much more extreme edge of parameter space.  To quantify the above statements, we follow \citetalias{diamond-stanic09} and \citet{wu12}, and we fit a lognormal profile to the \rew\ distributions of \civ, \mgii, and \hb\ for our comparison SDSS quasars.  The best-fit profiles have  mean $\log$~\rew=$1.59\pm0.19$, $1.48\pm0.14$, and $1.88\pm0.17$~\AA\ for \civ, \mgii, and \hb, respectively (error bars are standard deviations; the best-fit distributions for each species are similar to those found by \citealt{wu12}).\footnote{We quote the \rew[\mgii] distribution for the SDSS-IZ quasars, since those quasars have luminosities more similar to our X-shooter targets compared to the SDSS-LZ subset.  The \rew[\mgii] distribution for the SDSS-LZ subset is similar, however, with $\log$~\rew[\mgii]=$1.50\pm0.17$~\AA. Thus, there does not appear to be a large systematic bias when we compare to the SDSS-LZ subset, even though the SDSS-LZ quasars have systematically lower redshifts and luminosities.}  
As initially shown by \citetalias{diamond-stanic09} and \citet{wu12}, the SDSS quasar \rew[\civ] distribution shows a clear excess of quasars with \rew\ weaker than the $-$3$\sigma$ limit.  Of the 10956 SDSS-IZ quasars, 252 have \rew[\civ]$<$10.7~\AA\ ($<$$-$3$\sigma$), while only 33  have \rew[\civ]$>$141.3~\AA\ ($>$$+$3$\sigma$).    Neither \mgii\ or \hb\ display a similar excess of quasars in their $<$$-3\sigma$ tails.

The SDSS-LZ and SDSS-IZ samples  might contain a small number of WLQs in their low-\rew\ tails, even after removing objects identified as potential WLQs by \citet{collinge05} and \citetalias{plotkin10}.   Since we do not yet have physically motivated criteria for selecting WLQs, we do not attempt to exclude additional WLQs within the comparison samples, in order to avoid also removing relatively weak-lined but otherwise ``normal'' quasars  (see, e.g.,  Section 3.2 of \citealt{wu12} for a discussion on the (in)efficiency of WLQ selection).  We stress that any unidentified WLQs in the SDSS comparison  samples are too small in number to influence statistical conclusions.  For example, the log-normal fit to the \rew[\civ] distribution of SDSS-IZ quasars represents an accurate description of the ``normal'' quasars in that sample.  Furthermore, the presence of any unidentified WLQs in the comparison sample would only act to bias our WLQ and comparison samples to appear more similar, thereby strengthening our conclusions that WLQs have line properties on the weaker end of the parent quasar population.

\subsection{Rest-frame Optical Properties}
\label{sec:res:opt}
\begin{figure}
\includegraphics[scale=0.7]{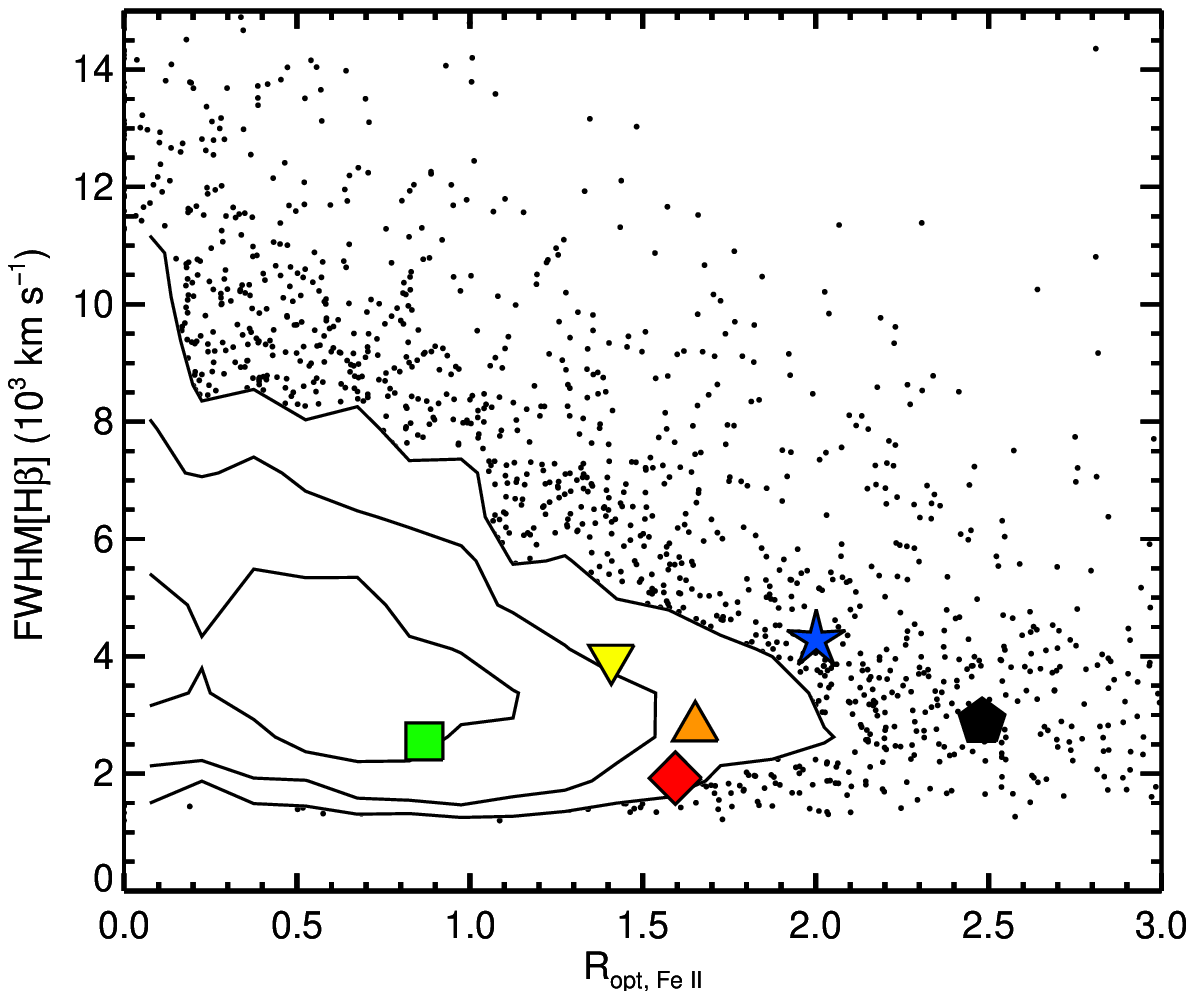}
\caption{FWHM of \hb\ versus \rfeiiopt, the rest-frame equivalent width ratio of optical \feii\ (4434--4684~\AA) to \hb, for the X-shooter WLQ targets (filled symbols) and the SDSS-LZ comparison quasars (contours and points). The colors and symbols shapes for the WLQs are the same as in Figure~\ref{fig:ewVew}; contours are logarithmically spaced by 0.35~dex, with the lowest contour denoting 20 quasars per bin ($20 \times 20$ bins along each axis).  Our X-shooter targets generally have enhanced  optical \feii\ and  FWHM[\hb]$\lesssim$4000~\kms, but they do not show rest-frame optical properties atypical from other quasars with enhanced optical \feii.}
\label{fig:ev1opt}
\end{figure}

Broad \ha\ and \hb\ emission lines are clearly detected in the NIR  spectra of all \nfit\ WLQs.  One source (\srcsix) has \rew[\hb]$\approx 16$~\AA, which is $4\sigma$ weaker than the mean of the SDSS-LZ sample; all other sources display \rew[\hb]$\approx$28-38~\AA, which is a rather typical value ($1.8-2.6\sigma$ lower than the mean).      Other WLQs with \hb\ coverage in the literature also display similar \rew[\hb], ranging from 20--50~\AA\ (see Table~\ref{tab:rew}).  All \nfit\ X-shooter targets  also show blended \feii\ emission near \hb.  Following \citet{boroson92}, we define the  ratio \rfeiiopt = \rew[\feii]$_{\rm opt}$/\rew[\hb], where \rew[\feii]$_{\rm opt}$ is measured from 4434--4684~\AA.  For all \nfit\ targets, \rfeiiopt\ lies toward the higher end of the expected distribution for the parent quasar distribution (i.e., \feii\ is  enhanced relative to \hb). That said, no WLQ shows an exceptionally large \rfeiiopt\  (see Figure~\ref{fig:ev1opt}).  WLQs generally possess rest-frame optical properties that are consistent with other quasars that display enhanced \feii\ emission.  For example, all \nfit\ X-shooter targets  have  FWHM[\hb] $\lesssim 4000$~\kms, which is  consistent with other quasars that have similar \rfeiiopt\ values  \citep[][also see Figure~\ref{fig:ev1opt}]{boroson92, sulentic00, sulentic07, shen14}.   

To quantify the above statements, we  compare the \rfeiiopt\ distribution of WLQs and SDSS-LZ quasars  by running Kolmogorov-Smirnov (K-S) and Mann-Whitney (M-W) tests.   The K-S and M-W tests are both non-parametric tests that assess if two (independent) distributions are drawn from the same parent population.  The K-S test achieves this by comparing the maximum deviation between the  cumulative distribution functions of the two observed populations, while the M-W test uses rank-ordering to compare the median values of the two distributions  \citep[e.g.,][]{sheskin11}.  Given the small number of objects in our WLQ sample, we opt to run both tests here as alternative ways to illustrate statistical conclusions.   The \rfeiiopt\ distribution of WLQs is different than that of the SDSS-LZ quasars at the $>$99\% level ($p=0.004$ from a K-S test; $p=0.001$ from a M-W test), indicating that WLQs as a population are indeed statistically more likely to display enhanced optical \feii\ emission.  We again stress that any potentially unidentified WLQs in the SDSS-LZ sample are too small in number to bias the above statistics (and any bias would only strengthen our conclusions, since the bias would force the two distributions to appear more similar).

We expect quasars with large \rfeiiopt\ to also show weak \oiii\ emission \citep[e.g.,][]{boroson92}.   Unfortunately, \oiii\ constraints are extremely limited for our X-shooter targets, as we only have \oiii\ coverage for the two targets at $z\approx1.4$ (i.e., \srcthree\ and \srcseven; the other four targets are at too high-redshift for 5007~\AA\ to fall within an atmospheric transmission window).   Indeed, only \srcthree\ with the weakest \feii\ emission (\rfeiiopt=0.9) has a firm \oiii\ detection, but constraints on more objects are required to determine if WLQ \oiii\ emission follows the trend expected from typical quasars.

\subsection{Rest-frame UV Properties}
\label{sec:res:uv}
The bluer sensitivity of the UVB arm compared to the SDSS provides our first constraints on \civ\ for two sources  (\srcthree\ and \srcseven, both are at $z\sim1.4$ and were included in the \citetalias{plotkin10} sample because of their small \rew[\mgii]).   These two sources display  the largest \rew[\civ]\ of all \nfit\ targets.  \srcthree\ has \rew[\civ]$\approx 19$~\AA, which is only 1.7$\sigma$ weaker than the mean \rew[\civ] of the SDSS-IZ comparison sample; \srcseven\ has \rew[\civ]$\approx 8$~\AA, which is 3.7$\sigma$ weaker than the mean.  Their stronger \civ\ therefore implies that weak \mgii\ does not  guarantee very weak \civ\ emission \citep[also see][for a similar conclusion]{wu12}.  It is noteworthy that both sources have \rew[\civ]$>$5~\AA, meaning that neither would  have passed the \citetalias{plotkin10} spectroscopic criteria for BL~Lac objects if SDSS covered \civ.

The UV emission lines  covered by each SDSS spectrum remain weak in each X-shooter observation.  For example, the four X-shooter targets with \civ\ coverage in SDSS still have \rew[\civ]$<$5~\AA.   We  also apply our fitting routine to the SDSS spectra to measure \rew[\mgii], and \mgii\ is not highly variable between the two epochs.\footnote{\citetalias{plotkin10} did not account for  Fe emission when they performed their spectral line measurements, because it is not necessary to do so for BL~Lac selection (since the beamed synchrotron jet emission  dilutes the Fe continuum).   As a result, \citetalias{plotkin10} systematically underestimate \rew[\mgii] when the continuum is not jet dominated, and we measure \rew[\mgii] greater than their 5~\AA\ threshold for all \nfit\ targets when accounting for Fe emission.  The \citetalias{plotkin10} measurements on \rew[\civ] are not affected by this systematic, because there is not a strong Fe continuum near \civ.} 
The fractional change between \rew[\mgii]  in the SDSS and X-shooter epochs are (\rew[\mgii]$_{\rm SDSS}$ - \rew[\mgii]$_{\rm Xshooter}$)/\rew[\mgii]$_{\rm Xshooter}$ = $0.41\pm0.18$, $0.05_{-0.03}^{+0.06}$, $0.34\pm0.29$, $-0.14_{-0.24}^{+0.15}$, $0.19\pm0.08$, $0.11\pm0.17$ for \srcone, \srctwo, \srcthree, \srcfive, \srcsix, and \srcseven, respectively.

The new systemic redshifts from the X-shooter data also allow improved insight into the rest-frame UV properties.  With updated redshifts, we can more rigorously investigate potential velocity offsets between the peaks of high- and low-ionization emission lines.  Five of our targets  (\srcone, \srctwo, \srcfive, \srcsix, \srcseven) display strong blueshifts in the peak wavelength of their \civ\ lines  ($\Delta v > 1000$~\kms; calculated relative to 1549~\AA), suggestive of a non-virialized component to the \civ\ emitting BELR  gas. With the new systemic redshifts from X-shooter, we find that two of these sources (\srctwo\ and \srcsix) also show significant \mgii\ blueshifts (calculated relative to 2800~\AA) as well ($1281\pm183$ and $624\pm180$~\kms, respectively; see Table~\ref{tab:linelum}).  These blueshifts are discussed  in Section~\ref{sec:disc:softsed}.  
 
\subsection{Black Hole Masses and Eddington Ratios}
\label{sec:res:bhmass}
We use the fits from the \hb\ complexes to estimate  virial black hole masses (\mbh) and  Eddington ratios (\lledd).  Since the optical properties from our X-shooter spectra are not atypical compared to normal quasars, we likely can obtain representative \mbh\ estimates from the \hb\ line (although see caveat below).   We use the  luminosity of the best-fit (linear) continuum at 5100~\AA, $L_{5100}$,  and the best-fit FWHM[\hb], along with the empirical BELR size-luminosity relation of \citet{kaspi05} (as modified by \citealt{bentz09}):

\begin{equation}
\frac{M_{\rm BH}}{10^6 M_{\odot}} = 5.05 \left[ \frac{L_{5100} }{10^{44}~{\rm erg~s^{-1}}}\right]^{0.5} \left[\frac{{\rm FWHM} \left( {\rm H}\beta\right)}{10^3~{\rm km~s}^{-1}}\right]^2 ,
\label{eq:mbh}
\end{equation}

\begin{equation}
L_{\rm bol}/L_{\rm Edd} = 0.13~f\left(L\right)  \left[ \frac{L_{5100}}{10^{44}~{\rm erg~s^{-1}}}\right]^{0.5} \left[\frac{{\rm FWHM} \left( {\rm H}\beta\right)}{10^3~{\rm km~s}^{-1}}\right]^{-2} ,
\label{eq:lledd}
\end{equation}

\noindent where $f(L)$ is a luminosity-dependent bolometric correction to $L_{5100}$.  We calculate $f(L)$ from Equation (21) of \citet{marconi04}.   We adopt the above relations for consistency with \citet{shemmer10} and \citet{wu11}, who presented \hb\ measurements for two  WLQs at $z\sim3.5$ and one \srcphl-analog at $z\sim2.2$, respectively.    The black hole mass and Eddington ratio estimates are listed in Table~\ref{tab:bhmass}.  We find black holes masses between $\log$(\mbh/\msun) $\approx 8-9$, and Eddington ratios \lledd$\approx 0.3-1.3$.   The  \lledd\ estimates are toward the high-end of the range typically observed for $z\sim1.4-1.7$ quasars, although we note that the uncertainties on each measurement are rather large (see Table~\ref{tab:bhmass}).

Our \nfit\ WLQs are at higher redshift  than  reverberation mapped quasars to which virial black hole masses are calibrated; our WLQs also  generally have larger \rfeiiopt\ and narrower \hb\ than most reverberation mapped quasars.  These caveats may introduce systematic biases into our \mbh\ and \lledd\  estimates \citep{richards11}.  We thus consider our \mbh\ and  \lledd\ estimates to be  approximate, but still useful for a qualitative analysis.   We do not attempt \mbh\ or \lledd\ estimates using either \civ\ or \mgii\ because  we have indications for non-virialized motion affecting those lines for some of our sources, which could bias the mass measurements \citep[see, e.g.,][]{baskin05, denney12, trakhtenbrot12, kratzer14}.

\section{Discussion}
\label{sec:disc}
We observed with X-shooter a sample of six SDSS WLQs at moderate-redshifts ($z=1.4-1.7$).  The extended spectral range provided by X-shooter compared to the SDSS allows a simultaneous comparison of the relative strengths of low- and high-ionization potential emission lines within individual objects.    The X-shooter spectra definitively illustrate that  when  high-ionization emission lines (i.e., \civ) are exceptionally weak,  lower-ionization lines (i.e., \hb) remain relatively normal (albeit on the weaker end of the distribution).     We do not see evidence for absorption being the cause of the weak high-ionization emission lines in any object, since we do not observe broad absorption troughs, and the power-law continua of our X-shooter spectra are typical of other Type~1 quasars (see Section \ref{sec:pl}; also see, e.g., \citetalias{diamond-stanic09}; \citealt{wu11, wu12}).  

\begin{figure}
\includegraphics[scale=0.7]{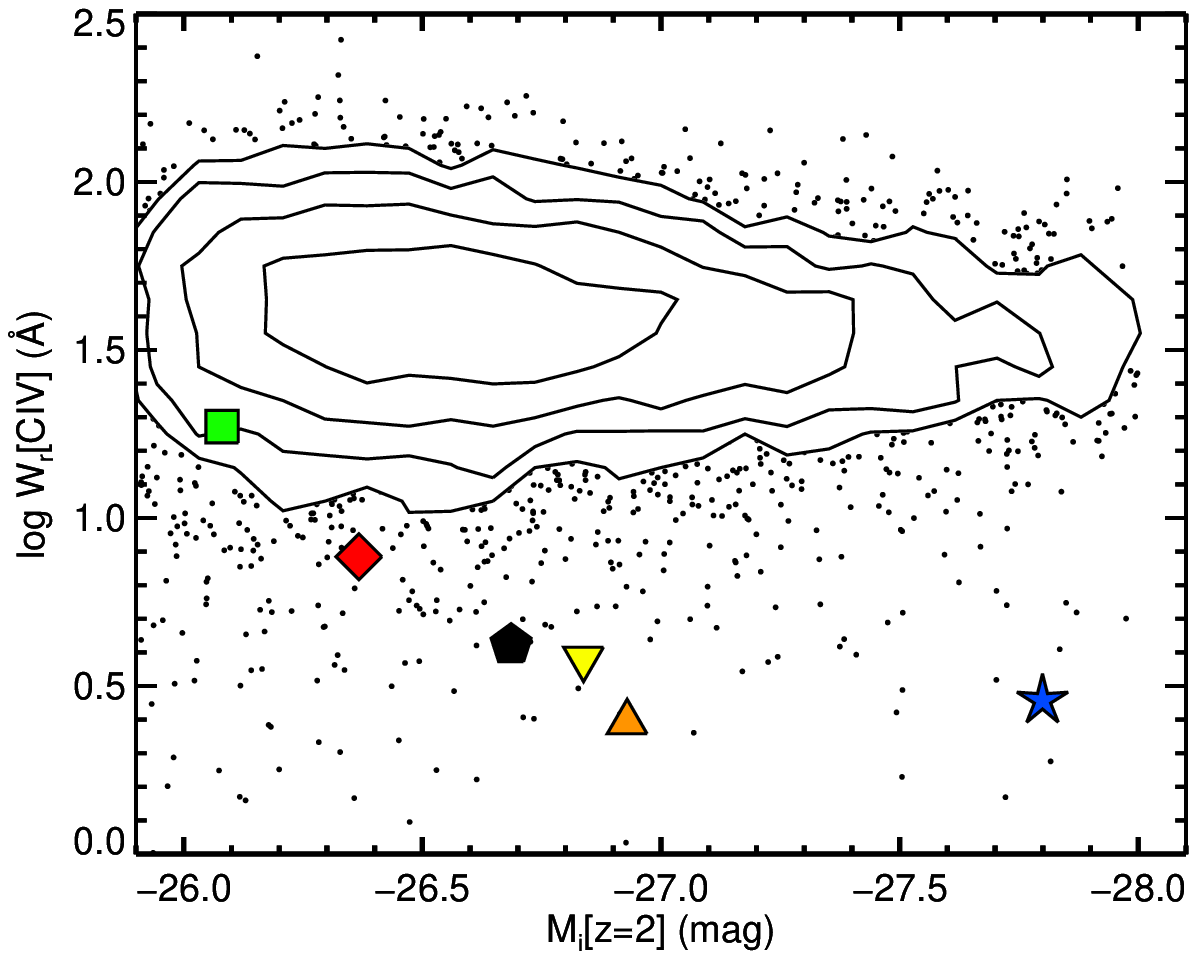}
\caption{Logarithm of the \civ\ rest-frame equivalent width versus $i$-band absolute magnitude for the X-shooter WLQ targets (filled symbols) and the SDSS-IZ comparison quasars (contours and points) All symbols have the same meaning as in Figure~\ref{fig:ewVew}.  The  $i$-band absolute magnitudes are K-corrected to $z=2$, as reported by  \citet{shen11}  for the SDSS-IZ  quasars and for our WLQs.  Except for \srcthree\ (and perhaps \srcseven), our WLQs have substantially weaker \civ\ rest-frame equivalent widths than expected given their continuum luminosities. }
\label{fig:baldwin}
\end{figure}

 As described below, the weak high-ionization lines are  unlikely to be driven by high-luminosity as might be expected from the Baldwin effect \citep{baldwin77}, at least for the four X-shooter targets with the smallest  \rew[\civ]$<$5~\AA.  The SDSS-IZ comparison quasars span similar luminosities as our X-shooter targets ($-28 < M_i < -26$~mag), and the SDSS-IZ quasars do not display a strong correlation between \rew[\civ] and $M_i$ over this relatively narrow luminosity range (Figure~\ref{fig:baldwin}); all four WLQs with \rew[\civ]$<$5~\AA\ in Figure~\ref{fig:baldwin} have weaker \civ\ than would be expected given their luminosities \citep[also see][]{shemmer15}.  Among the other two X-shooter targets, \srcthree\ does not have exceptionally weak \civ\ for its luminosity, while \srcseven\ is borderline.

\begin{figure}
\includegraphics[scale=0.7]{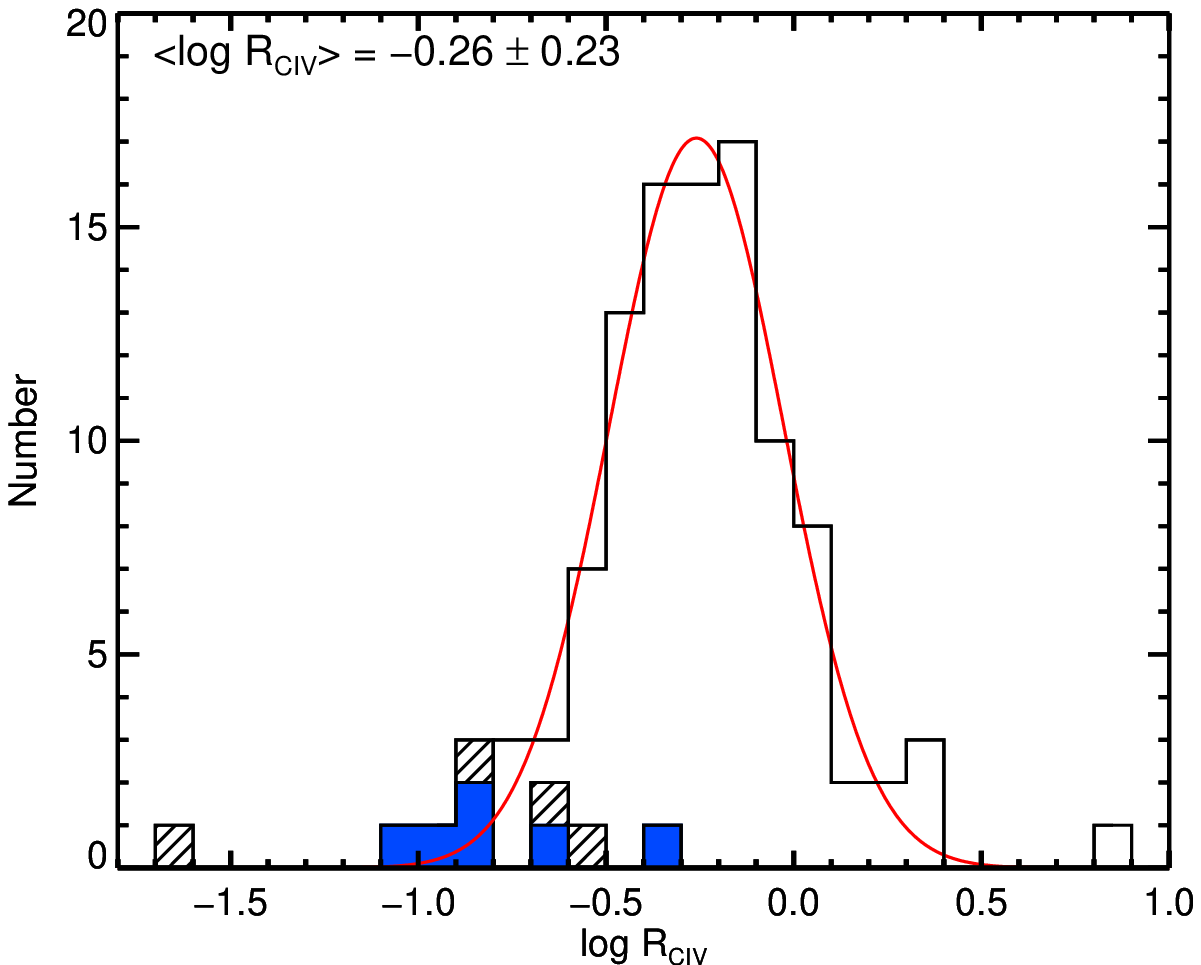}
\caption{Distribution of $\log R_{\rm CIV}= \log$(\rew[\civ]/\rew[\hb]) for the \nsmallqso\ object small quasar comparison sample (open histograms), with the best log-normal fit overdrawn (red curve).  Filled blue histograms show the $\log R_{\rm CIV}$ distribution for our \nfit\ X-shooter targets; hatched histograms show  other WLQs in the literature with \hb\ coverage (see Section~\ref{sec:res:otherwlqs}), excluding \srcpg\ which does not have an \hb\ detection ($\log~R_{\rm CIV}>-0.94$).   WLQs populate the small \rciv\ tail of the distribution, except for \srcthree\ which has a \rciv\ value typical of other quasars.}
\label{fig:rciv}
\end{figure}

To further illustrate the difference between high- and low-ionization lines, we examine the ratio  \rciv$=$\rew[\civ]/\rew[\hb].    The \nsmallqso\ quasars in our ``small'' comparison sample appear to follow a log-normal distribution with $\left<\log R_{\rm CIV}\right>=-0.26\pm0.23$ (quoted error is the standard deviation; see Figure~\ref{fig:rciv}).   The only X-shooter target with a typical \rciv\ ratio is \srcthree\ ($\log$\rciv$~\approx-0.30$).  This quasar also has a relatively normal \rew[\civ] and negligible \civ\ blueshift.   Therefore, \srcthree\  appears to simply fall on the weak-lined tail of the normal ``disk-dominated'' quasar population.  A relatively gas-deficient (but not completely anemic) BELR could provide an explanation for the relatively smaller (but not abnormal) \rew\ measures and the typical \rciv\ ratio for this source.  Considering the above, we do not  consider \srcthree\ to be a WLQ.

All of the other X-shooter targets have \rciv\ ratios toward the low-\rciv\ tail of the ``small'' quasar sample.  Their $\log$~\rciv\ values are not exceptionally small (i.e.,  1.5-3.4$\sigma$ weaker).  However, since our X-shooter targets were selected independent of their \hb\ properties, we would not expect all five to populate the lower end of the $\log$~\rciv\ distribution by random chance (i.e., randomly drawing  five sources that are $>$1.5$\sigma$ weaker than the mean of a log-normal distribution has  a probability $p\approx 10^{-6}$; a K-S test indicates that the WLQ \rciv\ distribution is different at the $p=0.003$ level, and a M-W tests finds $p=0.0002$).\footnote{The K-S and M-W tests are only meaningful if the \civ\ and \hb\ line properties are independent, such that selecting targets based on small \rew[\civ] does not bias \rew[\hb].  This assumption is likely only valid in certain circumstances, such as, e.g., the simplest ``anemic'' WLQ BELR scenarios.  There is little concern that the ``small'' comparison sample includes any unidentified WLQs.} 
 Of the  other WLQs/\srcphl-analogs with \hb\ coverage  in the literature (see Section~\ref{sec:res:otherwlqs}),  four have \hb\ detections.  These four sources   also display small \rciv\ values  (see hatched histograms in Figure~\ref{fig:rciv}), even though they also have relatively normal \rew[\hb] values  that are similar to our X-shooter targets (see Table~\ref{tab:rew}).  
 
In the simplest types of ``anemic'' BELR scenarios, the entire BELR would contain low amounts of gas.  In that case, we generally expect WLQs to have similar \rciv\ values as other quasars, and we would not expect to observe preferentially weakened (and blueshifted) high-ionization lines \citep[e.g.,][]{mcdowell95, dietrich02}. More complicated BELR physics must be invoked in order to explain why  WLQs have small \rciv\ ratios, in the context of anemic BELR scenarios.  Possible explanations include  a stratified BELR with a disrupted high-ionization component, or high- and low-ionization gas possessing different physical properties (e.g., covering factors, densities, etc.).  Regardless, it appears unlikely that low gas-content is the \textit{primary} cause of the WLQ phenomenon, although it certainly could play a secondary role.   A larger sample of NIR  spectra  of WLQs is needed to overcome small number statistics and probe the full continuum of rest-frame optical spectral properties, so that we can more rigorously  compare the distributions of line ratios between WLQs and the parent quasar population.

\subsection{Soft Ionizing Continua and Disk Winds}
\label{sec:disc:softsed}

If the WLQ phenomenon is not primarily driven by low gas content in the BELR, then the BELR is probably exposed to an unusually soft ionizing continuum.  The following  discusses soft ionizing continua in the context of ``disk/wind" models of the BELR \citep[e.g.,][]{murray95, elvis00, leighly04}.    In these models, lower-ionization potential lines (e.g., \mgii, \hb, \ha) are typically considered ``disk" lines, because they are associated with gas with kinematics  dominated by virialized motions \citep[e.g.,][]{eracleous03}.  High-ionization lines (like \civ) are thought of as ``wind" lines, because they can show a component with non-virialized motion,  often interpreted in the context of a radiatively line driven wind.  Some lines can have both a disk and wind component \citep[see, e.g.,][and references therein]{leighly04}.  In order to  drive a wind radiatively, the balance between the number of X-ray and UV photons is critical.  UV photons are required to accelerate the wind; however, a large X-ray flux will over-ionize the BELR gas, stripping too many valence electrons from the metals for the line-driving to be effective.   The above scenario is consistent with some well-known correlations between quasars, specifically relations between  X-ray properties and the relative balance between disk and wind line properties (i.e.,  the so-called Eigenvector 1 correlates; see, e.g., \citealt{boroson92, laor97, wills99, sulentic00, sulentic07, baskin05, shang07, gibson09, richards11, kruczek11, shen14}).

  \citet{wu11, wu12}  discuss in detail the relationship between rest-frame UV emission lines (in particular \civ\ blueshift) and X-ray properties for WLQs/\srcphl-analogs.   The ratio of X-ray luminosity to the luminosity of the rest-frame UV continuum provides constraints on the ionizing continuum (as does the relative strengths of some UV emission species), while  \civ\ blueshifts can be interpreted as a proxy for the strength of the disk wind (along the line of sight).   \citet{wu11,wu12} find that \srcphl-analogs (which were selected to have enhanced UV Fe emission, large \civ\ blueshifts,  and weak \rew[\civ]) are very X-ray weak.  On the other hand, WLQs that show weaker UV Fe emission (and typically also smaller \civ\ blueshifts) are generally X-ray normal.

\begin{figure}
\includegraphics[scale=0.7]{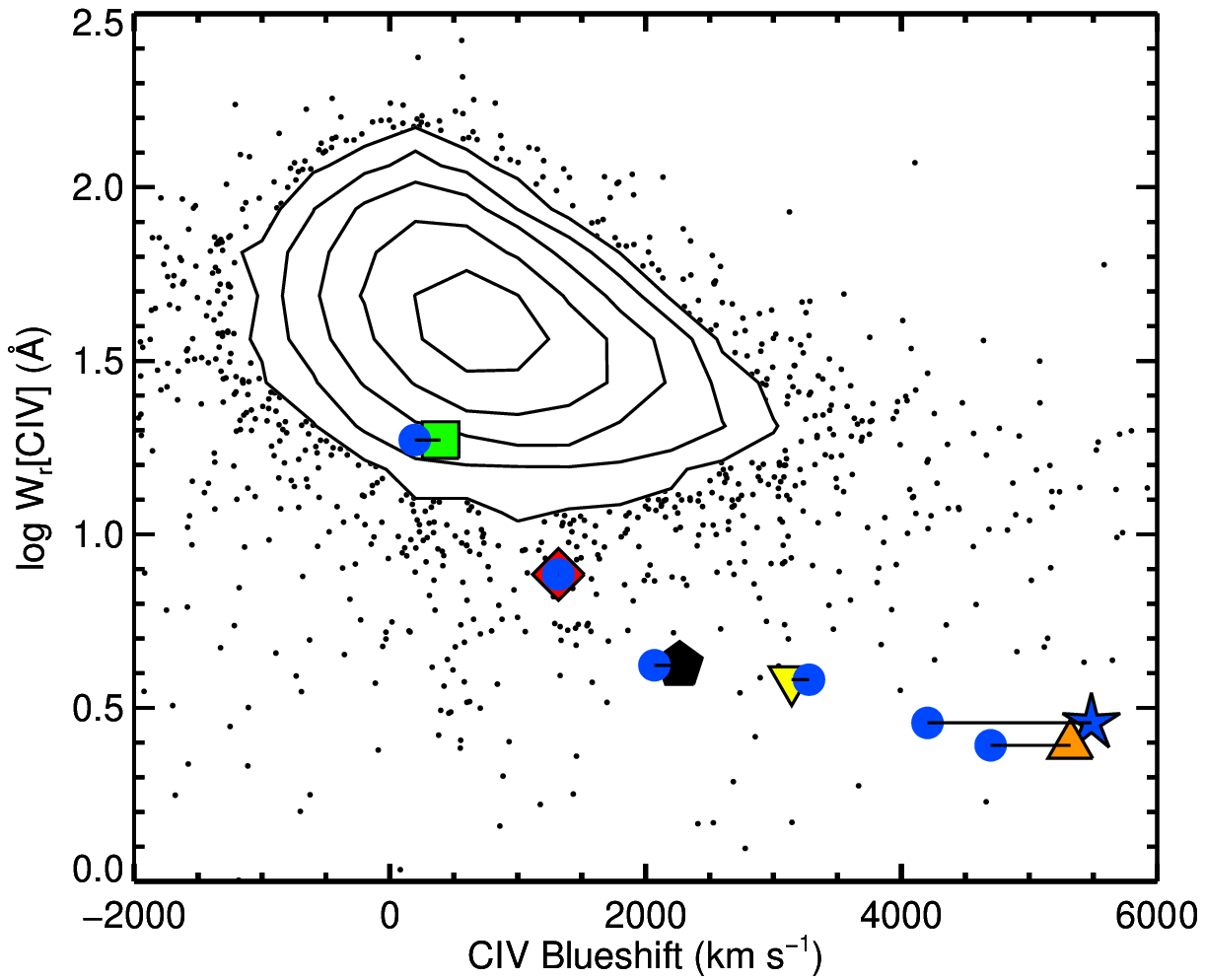}
\caption{Logarithm of \civ\ rest-frame equivalent width versus \civ\ blueshift (positive blueshifts denote line of sight motions moving toward the observer) for our X-shooter targets (filled symbols) and the SDSS-IZ sample (contours and small black circles).  The colors and symbol shapes for the WLQs are the same as in Figure~\ref{fig:ewVew}; contours are logarithmically spaced by 0.35~dex, with the lowest contour denoting 20 quasars ($20 \times 20$ bins along each axis).   For comparison to the SDSS-IZ sample, we plot the \civ\ blueshift relative to the \mgii\ line center for the X-shooter targets (filled blue circles), and also relative to the new systemic redshifts derived from the rest-frame optical spectra (other filled symbols; the symbols for each source are connected by a horizontal solid line to help guide the eye).   Our WLQs (except for \srcthree\ and perhaps \srcseven) appear to populate a distinct ``wind-dominated'' parameter space.}
\label{fig:ev1civ}
\end{figure}

 To examine the properties of our X-shooter targets in the context of the above phenomenological picture, we show our targets in the \civ\ \rew--blueshift plane in Figure~\ref{fig:ev1civ} (also see Figure~8 in \citealt{wu12}).  The comparison SDSS-IZ quasars in Figure~\ref{fig:ev1civ} have \civ\ blueshifts calculated relative to \mgii, so we present \civ\ blueshifts for our X-shooter targets  relative to both \mgii\ (blue circles) and to the X-shooter derived systemic redshift (other filled symbols).   \srcthree\ is the only target that clearly occupies a similar parameter space as typical quasars.  The other five sources lie in the ``wind-dominated'' quadrant of Figure~\ref{fig:ev1civ} \citep[see, e.g.,][]{richards11}, albeit with generally  much weaker \rew[\civ] than other quasars displaying similar blueshifts.\footnote{There is likely a link between the  FWHM[\hb]--\rfeiiopt\  plane (Figure~\ref{fig:ev1opt})  and the  \civ\ \rew--blueshift plane (Figure~\ref{fig:ev1civ}), in that quasars that fall in one quadrant in one plane typically map to a similar quadrant in the other plane \citep[e.g.,][]{sulentic07, richards11}.     Indeed,  the \nsoft\ X-shooter targets with large  \civ\ blueshifts ($>$1000~\kms) also have rest-frame optical properties typical of wind-dominated quasars (i.e.,  large \rfeiiopt\ and narrower FWHM[\hb]).}

\begin{figure}
\includegraphics[scale=0.7]{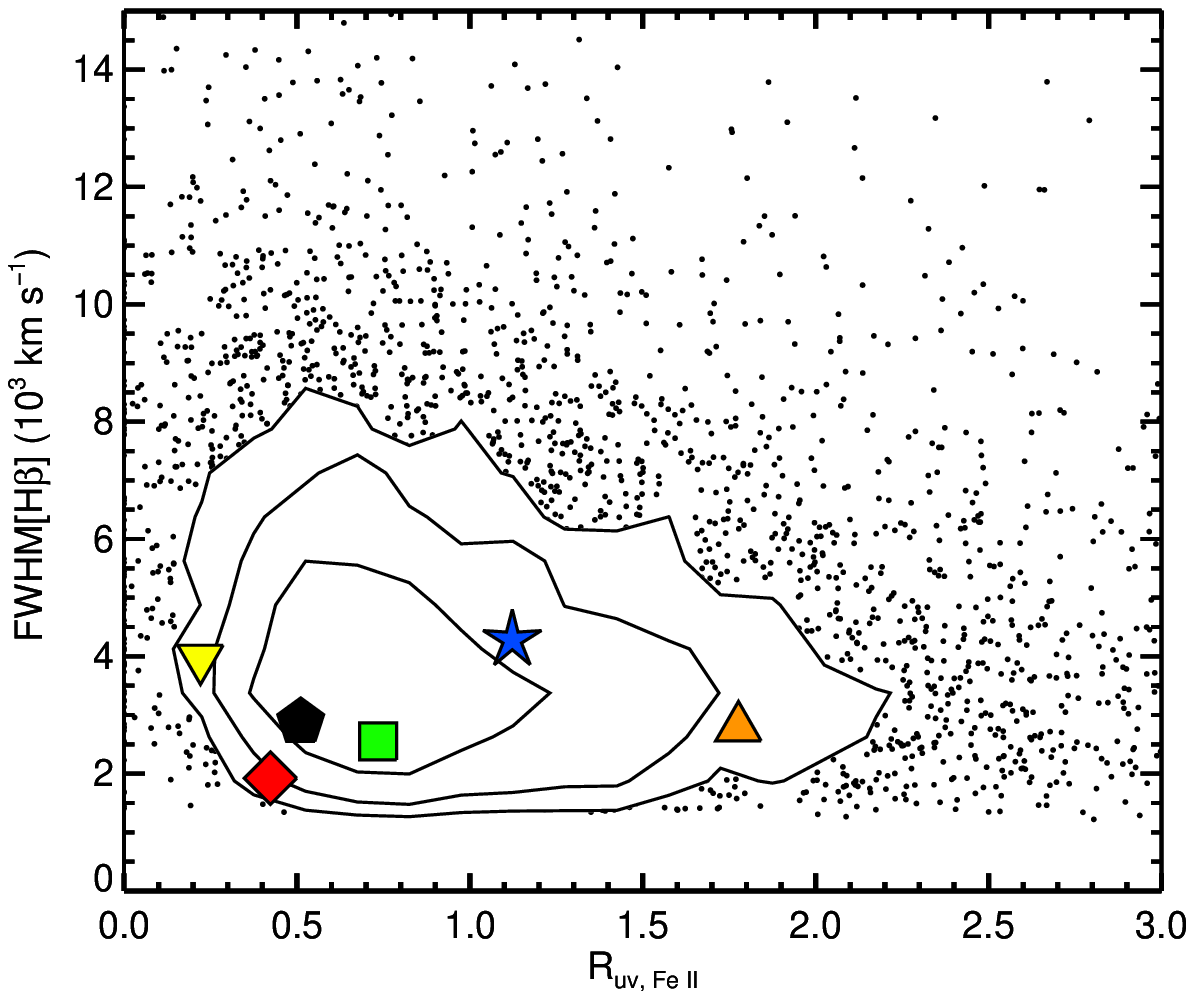}
\caption{Similar to Figure \ref{fig:ev1opt}, except that the x-axis here shows the rest-frame equivalent width ratio of UV \feii\ (2250--2650~\AA) to \hb, compared to the SDSS-LZ quasars (all symbols have the same meaning as in Figure~\ref{fig:ev1opt}).  Only the two WLQs with the strongest \civ\ winds (\srctwo\ and \srcsix) show enhanced UV \feii\ emission. }
\label{fig:ev1uv}
\end{figure}

  It is interesting that the two X-shooter targets with the strongest \civ\ blueshifts are the only two that also display blueshifted \mgii\ emission.  \mgii\ is often thought of as a disk emission line, given its relatively low ionization potential ($\chi_{\rm ion}\approx8$~eV).  However, \mgii\ can still have a non-negligible cross section to UV radiation, and it  sometimes has a wind component \citep[e.g.,][]{shang07}.   From our X-shooter sample, it appears that a relatively strong wind  is required to cause \mgii\ to display an outflowing component.  We also speculate that there could be a connection between the wind strength and the amount of \feii\ UV emission.  We quantify the strength of the \feii\ UV emission as   \rfeiiuv=\rew[\feii]$_{\rm uv}$/\rew[\hb], where \rew[\feii]$_{\rm uv}$ is calculated from 2250--2650~\AA, which we show relative to FWHM[\hb] in Figure~\ref{fig:ev1uv}.   Similar to the rest-frame optical \feii\ emission (i.e., \rfeiiopt; Figure~\ref{fig:ev1opt}), none of the X-shooter targets displays abnormally large \rfeiiuv.   However,  the two WLQs showing \mgii\ blueshifts   are the only ones to display relatively enhanced \rfeiiuv\ (compared to all five sources with \civ\ blueshifts  showing enhanced \rfeiiopt).    \mgii\ blueshift is therefore a newly identified parameter that could ultimately provide additional clues on the BELR properties of WLQs (also see \citealt{luo15} for a discussion on UV \feii\ and X-ray weakness).

\subsubsection{Weak \heii\ Emission}
\label{sec:disc:softsed:heii}
The strengths of the \heii\ lines at 1640 and 4686~\AA\ reflect the strength of the far-UV continuum at $\gtrsim$54~eV\ (i.e.,  photon energies required to produce singly ionized He), and therefore provide a means to test the soft SED hypothesis.  A large \rew[\heii]  implies a strong continuum that would overionize the BELR, thereby inhibiting the formation of strong winds (an anti-correlation between \rew[\heii] and disk wind velocity has indeed been observed by several studies; e.g., \citealt{leighly04a, richards11, baskin13, bowler14}). 

All \nfit\ X-shooter spectra cover the   \uvheii\ line, and four spectra (excluding \srcthree\ and \srcseven)  cover the \optheii\ line in the NIR arm.  We measure \rew\ values for each \heii\ line, with the $\lambda$1640 line measured between 1620-1650~\AA\ (which we refer to as \rewuvheii), and the $\lambda$4686 line measured between 4665-4700~\AA\ (\rewoptheii).   The UV and optical measurements are performed relative to the linear continua obtained from fitting the \civ\ and \hb\ complexes in Section~\ref{sec:fits}, respectively (and we remove the best-fit \feii\ continuum prior to measuring \rewoptheii).

All \heii\ lines are weak and measured to have \rew$<$1~\AA.  To assess the statistical significance of any potential \heii\ emission, and to assign meaningful limits on non-detections, we run Monte Carlo simulations as follows.  We make the null assumption that no \heii\ line is present in any spectrum.  We then take the best-fit spectral model for each \civ\ and \hb\ complex, to which we  add simulated (statistical) noise, such that each simulated spectrum follows $f_{\lambda, \rm sim} = f_{\lambda, \rm model} \pm f_{\lambda, \sigma}$.  At each wavelength $\lambda$, $f_{\lambda, \rm sim}$ is the simulated flux density, $f_{\lambda, \rm model}$ is the flux density from the best-fit model, and $f_{\lambda, \sigma}$ is randomly drawn from a Gaussian distribution with mean zero and a standard deviation set to the  error bar on each observed flux density (these error bars were calculated during the initial data reduction in Section~\ref{sec:obs:datared}).  We create 1000 simulated spectra for each source and complex, and we measure \rew[\heii] in each simulated spectrum.  We take the standard deviation of each $N=1000$ distribution of \rew\ measures ($\pm\sigma_{\rm rms}$) to represent the 68\% confidence interval for measuring a given \rew, in the case where variations in \rew\ are due solely to statistical fluctuations.  As expected, all simulated \rew\ distributions appear to follow a normal distribution with $\langle$\rew$\rangle \approx 0$.  

To consider \heii\ to be detected in emission in our X-shooter spectra, we require \rew[\heii]$> 3\sigma_{\rm rms}$.  Measured values for \rewuvheii\ and \rewoptheii\ are listed in Table~\ref{tab:rew} (for non-detections, we list the 3$\sigma_{\rm rms}$ upper-limit).   In the UV, only \srcthree\ and \srcseven\ are clearly detected, and  \srcsix\ is marginally detected.    In the optical, none of the four sources with spectral coverage display detectable \optheii.  Given the poorer $S/N$ in the X-shooter NIR arm,  our limits on \optheii\ are not highly constraining.  For ``normal'' non-BAL quasars with similar luminosities as our X-shooter targets, we expect \rewoptheii$\lesssim$5~\AA\ \citep[][a typical value may even be as small as 0.7~\AA, according to \citealt{vanden-berk01}]{dietrich02}, while most of our X-shooter targets are only constrained to  have \rewoptheii$<$7-8~\AA.

In the UV, our constraints on \uvheii\ are  meaningful (except for \srcfive), and generally consistent with the idea that WLQs have soft ionizing continua.   \citet{bowler14} find that  a composite spectrum of $\sim$1000 SDSS quasars at similar redshift and luminosity as our X-shooter targets has \rewuvheii$\approx 0.6$.  Only two of our X-shooter targets have a larger \rewuvheii.  However, these two objects (\srcthree\ and \srcseven) are the least exotic sources within our sample, and they unlikely lie on as extreme of an edge of parameter space as the other targets: we argued earlier that \srcthree\ is likely a relatively weak-lined but normal ``disk-dominated'' quasar; we discuss in Section~\ref{sec:disc:1447} that \srcseven\ may lie on the weak-lined tail of the normal ``wind-dominated'' quasar population.  All three of  the other X-shooter targets with meaningful limits have \rewuvheii$<$0.6.    As described earlier, these sources  also display strong \civ\ blueshifts, which  supports the notion that  WLQs are exotic versions of ``wind-dominated' quasars with softer than typical ionizing continua.

\subsubsection{Is the Ionizing Continuum Intrinsically Soft?} 
\label{sec:disc:bigbh}

The relatively normal rest-frame optical properties  suggest that WLQ BELRs are in an unusual ionization state, likely as a result of  a  soft ionizing continuum.   Here, we  exclude the possibility of an intrinsically soft SED due to a very massive SMBH.  \citet{laor11} show that there is a critical mass of  $M_{\rm BH} > 3.6\times10^{9}$~\msun\ for a non-spinning black hole ($a=0$), or  $> 1.4\times10^{10}$~\msun\ for a nearly maximally spinning prograde black hole ($a=0.998$), where the accretion disk becomes too cold to emit enough high-energy photons to ionize the BELR.  Taking our \hb-derived black hole masses at face value, no X-shooter target (or any other WLQ with \hb\ coverage) has a black hole mass  this large (although see caveats in Section~\ref{sec:res:bhmass}).  \citet{laor11} also predict that \hb\ should be very weak and broad (FWHM[\hb]$>$8000 and 16000~\kms\ for $a=0$ and 0.998, respectively), which is also not observed for any WLQ.  Finally, \citet{laor11} predict that  a cold accretion disk should peak at a wavelength $\lambda \gtrsim 3200$~\AA, which would imply that the continuum in the UVB arm of our X-shooter spectra should be redder  than is observed (i.e., $\alpha_{\lambda}$ should be larger or even positive; X-shooter places stronger constraints on the continuum shape than SDSS because of its bluer sensitivity).   Although very massive SMBHs are unlikely to be the explanation for our X-shooter targets, we stress that we cannot exclude the possibility that it is responsbile for weak lines in a handful of exceptional quasars (e.g., \citealt{laor11} do not maintain that it must apply to every single WLQ, explicitly noting that it is unlikely to be the explanation for  \srcpg).\footnote{Recent studies suggest that luminous quasars at $z\sim1.5-2$ are associated with high SMBH spins \citep{netzer14, trakhtenbrot14}.  This raises a possibility that some rare, potentially weak-lined objects in SDSS may represent a missing population of low-spin SMBHs.   From our X-shooter results, WLQ black hole masses appear to be too small for WLQs to represent such a population.}

The other two possibilities within the ``soft continuum''  group of WLQ models are an intrinsically soft SED due to super-Eddington accretion \citep{leighly07a, leighly07}, and an SED that is modified by  X-ray shielding gas before it illuminates the BELR \citep{wu11, wu12}.   If super-Eddington accretion is always associated with quenched X-ray emission, then the super-Eddington model cannot explain X-ray normal WLQs (about half of the population; \citealt{wu12}).  Among the X-ray weak WLQs, however, we cannot gain substantially new insight from the limited X-shooter sample into whether the SED is intrinsically soft or modified by shielding gas.  Further distinguishing between the two scenarios will require  statistically meaningful samples of WLQs with rest-frame optical spectral coverage, in parallel to ongoing efforts to observationally constrain the broadband properties \citep[e.g.,][]{luo15}.    For example, one  expects the ratio of high- vs.\ low-ionization species (e.g., \rciv)  to correlate with SED shape in the super-Eddington scenario.    On the other hand, high- vs.\ low-ionization line ratios should be less sensitive to the observed SED in the shielding gas model (since the observed X-ray weakness is  determined primarily by orientation).

 \subsubsection{\srcseven}
 \label{sec:disc:1447}
   \srcseven\  has only a modestly small  \rew[\civ]$\approx$8~\AA\ and a relatively slow disk wind compared to the other X-shooter targets (\civ\ blueshift $\approx$1300~\kms).   It is unclear whether or not this source is a very exotic quasar.   For example,  although its \rew[\civ] is  $\sim$4$\sigma$ weaker than the mean of the SDSS-IZ quasar distribution,  its location in Figure~\ref{fig:ev1civ} is not exceptionally different than other  quasars with similar \civ\ blueshifts.  We expect higher Eddington  ratio quasars to display smaller \rew[\civ] (i.e., the ``modified Baldwin Effect''; \citealt{baskin04}), and the \rew[\civ] of  \srcseven\ is  only $\sim$1.5$\sigma$ weaker than expected for its \lledd\ \citep{shemmer15}.  There is likely overlap between the weak-lined tail of the normal (wind-dominated) quasar population with relatively high-\lledd\ \citep[e.g.,][]{baskin04, netzer04, shen14}, and between the strong-lined tail of the WLQ population.   \srcseven\ seems to display properties in this grey area (see \citealt{shemmer15} for further discussion). 

\subsection{\citet{hryniewicz10}}
\label{sec:anemic:adolescent} 
There is a final possibility that could explain the relatively normal \hb\ and the weak/blueshifted \civ\ in our X-shooter spectra.   \citet{hryniewicz10} propose a model where WLQs represent a short-lived phase ($\sim$10$^3$ yr) of an evolutionary sequence where black hole activity has just recently been (re)activated.  They propose that WLQs are essentially ``adolescent'' quasars, where the BELR has not yet finished forming.   Low-ionization  species (like \hb\ and \mgii) that form close to the accretion disk should  appear normal.  However, the disk wind has not yet had sufficient time to populate regions farther above the disk, and higher-ionization ``wind'' species like \civ\ could appear exceptionally weak.  We cannot effectively discriminate between adolescent quasars and soft ionizing continua from this X-shooter program alone, because each scenario predicts similar spectral properties.  Testing this scenario would require stringent constraints on WLQ evolution (i.e., more WLQs would be expected at higher redshifts) and/or the WLQ luminosity function.  There are intriguing indications that WLQs are indeed more common at higher-redshifts, from studies on the fraction of WLQs at $z\approx3-6$ (\citetalias{diamond-stanic09}; \citealt{banados14}).  The systematics involved in WLQ selection at lower-redshifts are not  understood well enough to extend this type of work to $z<3$ at the moment.  However, such systematics are starting to be addressed,  between this X-shooter program and other NIR spectroscopic campaigns on  $z<3$ SDSS WLQs \citep[e.g.,][]{shemmer10, wu11}, and also the recent efforts by \citet{meusinger14} at WLQ selection (we note, however, that the the \citetalias{plotkin10} and \citealt{meusinger14} WLQ samples likely suffer from different sets of selection-based systematics, and as a result may not recover identical populations of rare quasars).

\subsection{Arguments Against Synchrotron Beaming}
\citet{meusinger14} recently applied machine learning techniques to the SDSS spectroscopic database to assemble a new WLQ sample, and they find that a substantial fraction of their WLQs have radio-detections in FIRST (approximately one-quarter of their WLQ sample are radio-loud).  In light of this work, and the fact that \srcpg\ may also have a beamed (albeit weak) relativistic jet \citep{blundell03, gallo06}, we revisit the prospect of relativistic beaming here.  A beamed  jet is still highly unlikely to cause the weak BELRs for our X-shooter targets (as well as for \srcpg), for the following reasons.  First, all X-shooter targets were required to be radio-fainter, at the $>$3$\sigma$ level, than the population of radio-loud SDSS BL~Lac objects in \citetalias{plotkin10}.  Therefore, any radio jet must be low luminosity and/or weakly beamed.   Any beamed continuum would also be highly variable, so that we would expect variable \rew\ measures between the SDSS and X-shooter epochs \citep[e.g.,][]{ruan14}, which we do not observe.  Furthermore, dilution from a relativistic jet would reduce the \rew\ of all lines, inconsistent with our observations of enhanced optical \feii\ (i.e., large \rfeiiopt).  Finally, the \civ\ blueshifts cannot be explained by dilution from a beamed jet, and the magnitude of the blueshifts  (i.e., $\sim$10$^3$~\kms) for our X-shooter targets are  uncommon for radio-loud quasars \citep{richards11}.

\section{Conclusions}

We presented X-shooter spectroscopy  for  \nfit\ WLQs at moderate redshift ($z=1.4-1.7$), allowing us to compare the properties of high- vs.\ low-ionization emission species in individual objects.   The NIR spectra from X-shooter more than double the number of WLQs with spectral coverage of the \hb\ complex.   Although all of our WLQs  display relatively weak \hb\ emission,  none has a rest-frame optical spectrum that is  abnormal compared to other quasars. With new constraints on \civ\ for two targets (which were lacking from SDSS), one target (\srcthree; $z=1.42$) appears to be a relatively weak-lined but otherwise normal ``disk-dominated'' quasar.  The other five targets have relatively narrow \hb\ and slightly enhanced optical Fe emission, which is typical for ``wind-dominated'' quasars.  Indeed, all five display \civ\ blueshifts $>$1000~\kms.  One of these quasars (\srcseven; $z=1.43$) may simply lie on the weak-lined tail of the normal, high-\lledd\ ``wind-dominated'' quasar population.  The other four targets, however, have exceptionally weak \rew[\civ]$<$5~\AA\ (and apparently weak \uvheii), and they appear to be bona fide WLQs.    Exceptionally weak \civ\ emission therefore does not guarantee very weak emission from low-ionization lines (e.g., \hb), which is consistent with \citet{wu12}, who concluded that it is extraordinarily difficult to select WLQs based solely on their low-ionization emission lines.  In other words, WLQs appear to separate most cleanly from  the ``normal'' quasar population via their high-ionization lines, especially \civ\ and \lyanv.  

As we are now starting to build a respectably sized sample of WLQs with rest-frame optical spectra (six from this work, and five others in the literature), we determine that low amounts of gas in the BELR is unlikely the primary cause of the WLQ phenomenon.  WLQs have relatively small \rciv=\rew[\civ]/\rew[\hb] line ratios, which is inconsistent with their BELRs simply having low gas content or small covering factors.  Instead, either high-ionization lines have different physical properties than low-ionization lines, or WLQ BELRs are  in an unusual photoionization state due to a soft SED.  In the latter case, a larger sample of WLQs with rest-frame optical spectra are required to determine if the SED is intrinsically soft (as predicted if WLQs are super-Eddington accretors; \citealt{leighly07a, leighly07}) or if the SED is modified prior to illuminating the BELR \citep{wu11}.   Our X-shooter targets are unlikely to have soft SEDs due to very massive black holes \citep[e.g.,][]{laor11}.   

There is likely a connection between the shape of the ionizing continuum and the strength of a disk wind, and WLQs indeed appear to be exotic versions of ``wind-dominated'' quasars.  \citet{wu11}  show that WLQs with the largest \civ\ blueshifts (i.e., strong winds) also tend to  be X-ray weak, and they typically show strong UV Fe emission.  Our two X-shooter targets  with the largest \civ\ blueshifts (\srctwo\ and \srcsix) are also the only two that show enhanced UV Fe emission, and they are the only two to show significant \mgii\ blueshifts.   We thus add a new piece to the above picture, in that \mgii\ blueshift could be an additional parameter that might eventually help provide  insight into WLQs.

Finally, we note that it is entirely reasonable that WLQs could be formed through multiple channels, and the WLQ population could very likely represent a heterogeneous population.  More than one mechanism could contribute within individual objects as well (e.g., a single quasar could have weak lines due to a combination of a very massive black hole accreting at a super-Eddington rate, with excess shielding gas).  The ultimate goal is therefore to determine if one mechanism tends to dominate over others, or if multiple mechanisms contribute at comparable levels (which could help us learn about the central engine in other types of less extreme quasars).  Even with the small sample size considered in this study, we can already exclude low  gas content across the entire BELR as the primary mechanism.  We also identify a potential source of contamination to the \citet{collinge05} and \citetalias{plotkin10} WLQ samples, in that some $z \approx 1.4$ SDSS WLQs may simply lie on the weak-lined tail of the normal quasar population (but not be exotic quasars in the same sense as other WLQs).  These observations provide useful insight to assist in identifying lower-redshift WLQs  in the future, in order to eventually build representative distributions of the line properties (e.g., \rciv) to compare to X-ray properties of the WLQ population.  This information will also be critical to constrain WLQ evolution, in order to use lower-redshift WLQs to further investigate indications that WLQs are more common at higher redshift \citep[e.g.,][]{hryniewicz10, liu11, banados14}.  Understanding the physical nature of WLQs will ultimately provide new insight into the properties of quasar BELRs in general.

\acknowledgements
We thank the anonymous referee for constructive comments that improved this paper.  We thank Lucas Ellebroek and Hughes Sana for advice on reducing the X-shooter spectra.  WNB and BL acknowledge support from  NASA ADP grant NNX10AC99G, and  Chandra X-ray Center grant GO3-14100X.   XF acknowledges support from NSF grant AST 11-07862.  This research has made use of the NASA/IPAC Extragalactic Database (NED) which is operated by the Jet Propulsion Laboratory, California Institute of Technology, under contract with the National Aeronautics and Space Administration.  This research has also made use of the SIMBAD database, operated at CDS, Strasbourg, France


\begin{thebibliography}{89}
\expandafter\ifx\csname natexlab\endcsname\relax\def\natexlab#1{#1}\fi

\bibitem[{{Anderson} {et~al.}(2001){Anderson}, {Fan}, {Richards}, {Schneider},
  {Strauss}, {Vanden Berk}, {Gunn}, {Knapp}, {Schlegel}, {Voges}, {Yanny},
  {Bahcall}, {Bernardi}, {Brinkmann}, {Brunner}, {Csab{\'a}i}, {Doi},
  {Fukugita}, {Hennessy}, {Ivezi{\'c}}, {Kunszt}, {Lamb}, {Loveday}, {Lupton},
  {McKay}, {Munn}, {Nichol}, {Szokoly}, \& {York}}]{anderson01}
{Anderson}, S.~F., {Fan}, X., {Richards}, G.~T., {et~al.} 2001, \aj, 122, 503

\bibitem[{{Ba{\~n}ados} {et~al.}(2014){Ba{\~n}ados}, {Venemans}, {Morganson},
  {Decarli}, {Walter}, {Chambers}, {Rix}, {Farina}, {Fan}, {Jiang}, {McGreer},
  {De Rosa}, {Simcoe}, {Wei{\ss}}, {Price}, {Morgan}, {Burgett}, {Greiner},
  {Kaiser}, {Kudritzki}, {Magnier}, {Metcalfe}, {Stubbs}, {Sweeney}, {Tonry},
  {Wainscoat}, \& {Waters}}]{banados14}
{Ba{\~n}ados}, E., {Venemans}, B.~P., {Morganson}, E., {et~al.} 2014, \aj, 148,
  14

\bibitem[{{Baldwin}(1977)}]{baldwin77}
{Baldwin}, J.~A. 1977, \apj, 214, 679

\bibitem[{{Baskin} \& {Laor}(2004)}]{baskin04}
{Baskin}, A., \& {Laor}, A. 2004, \mnras, 350, L31

\bibitem[{{Baskin} \& {Laor}(2005)}]{baskin05}
---. 2005, \mnras, 356, 1029

\bibitem[Baskin et al.(2013)]{baskin13} Baskin, A., Laor, A., 
\& Hamann, F.\ 2013, \mnras, 432, 1525 

\bibitem[{{Bentz} {et~al.}(2009){Bentz}, {Peterson}, {Netzer}, {Pogge}, \&
  {Vestergaard}}]{bentz09}
{Bentz}, M.~C., {Peterson}, B.~M., {Netzer}, H., {Pogge}, R.~W., \&
  {Vestergaard}, M. 2009, \apj, 697, 160

\bibitem[{{Blandford} \& {Rees}(1978)}]{blandford78}
{Blandford}, R.~D., \& {Rees}, M.~J. 1978, in BL Lac Objects, ed. A.~M.
  {Wolfe}, 328--341

\bibitem[{{Blundell} {et~al.}(2003){Blundell}, {Beasley}, \&
  {Bicknell}}]{blundell03}
{Blundell}, K.~M., {Beasley}, A.~J., \& {Bicknell}, G.~V. 2003, \apjl, 591,
  L103

\bibitem[{{Boroson} \& {Green}(1992)}]{boroson92}
{Boroson}, T.~A., \& {Green}, R.~F. 1992, \apjs, 80, 109

\bibitem[Bowler et al.(2014)]{bowler14} Bowler, R.~A.~A., 
Hewett, P.~C., Allen, J.~T., \& Ferland, G.~J.\ 2014, \mnras, 445, 359 

\bibitem[{{Cardelli} {et~al.}(1989){Cardelli}, {Clayton}, \&
  {Mathis}}]{cardelli89}
{Cardelli}, J.~A., {Clayton}, G.~C., \& {Mathis}, J.~S. 1989, \apj, 345, 245

\bibitem[{{Collinge} {et~al.}(2005){Collinge}, {Strauss}, {Hall}, {Ivezi{\'c}},
  {Munn}, {Schlegel}, {Zakamska}, {Anderson}, {Harris}, {Richards},
  {Schneider}, {Voges}, {York}, {Margon}, \& {Brinkmann}}]{collinge05}
{Collinge}, M.~J., {Strauss}, M.~A., {Hall}, P.~B., {et~al.} 2005, \aj, 129,
  2542

\bibitem[{{Czerny} {et~al.}(2011){Czerny}, {Hryniewicz}, {Niko{\l}ajuk}, \&
  {S{\c a}dowski}}]{czerny11a}
{Czerny}, B., {Hryniewicz}, K., {Niko{\l}ajuk}, M., \& {S{\c a}dowski}, A.
  2011, \mnras, 415, 2942

\bibitem[{{Denney}(2012)}]{denney12}
{Denney}, K.~D. 2012, \apj, 759, 44

\bibitem[{{Diamond-Stanic} {et~al.}(2009){Diamond-Stanic}, {Fan}, {Brandt},
  {Shemmer}, {Strauss}, {Anderson}, {Carilli}, {Gibson}, {Jiang}, {Kim},
  {Richards}, {Schmidt}, {Schneider}, {Shen}, {Smith}, {Vestergaard}, \&
  {Young}}]{diamond-stanic09}
{Diamond-Stanic}, A.~M., {Fan}, X., {Brandt}, W.~N., {et~al.} 2009, \apj, 699,
  782 (DS09)

\bibitem[{{Dietrich} {et~al.}(2002){Dietrich}, {Hamann}, {Shields},
  {Constantin}, {Vestergaard}, {Chaffee}, {Foltz}, \&
  {Junkkarinen}}]{dietrich02}
{Dietrich}, M., {Hamann}, F., {Shields}, J.~C., {et~al.} 2002, \apj, 581, 912

\bibitem[{{Elitzur} \& {Ho}(2009)}]{elitzur09}
{Elitzur}, M., \& {Ho}, L.~C. 2009, \apjl, 701, L91

\bibitem[{{Elvis}(2000)}]{elvis00}
{Elvis}, M. 2000, \apj, 545, 63

\bibitem[{{Eracleous} \& {Halpern}(2003)}]{eracleous03}
{Eracleous}, M., \& {Halpern}, J.~P. 2003, \apj, 599, 886

\bibitem[{{Fan} {et~al.}(1999){Fan}, {Strauss}, {Gunn}, {Lupton}, {Carilli},
  {Rupen}, {Schmidt}, {Moustakas}, {Davis}, {Annis}, {Bahcall}, {Brinkmann},
  {Brunner}, {Csabai}, {Doi}, {Fukugita}, {Heckman}, {Hennessy}, {Hindsley},
  {Ivezi{\'c} }, {Knapp}, {Lamb}, {Munn}, {Pauls}, {Pier}, {Rockosi},
  {Schneider}, {Szalay}, {Tucker}, \& {York}}]{fan99}
{Fan}, X., {Strauss}, M.~A., {Gunn}, J.~E., {et~al.} 1999, \apjl, 526, L57

\bibitem[{{Gallo}(2006)}]{gallo06}
{Gallo}, L.~C. 2006, \mnras, 365, 960

\bibitem[{{Gibson} {et~al.}(2009){Gibson}, {Jiang}, {Brandt}, {Hall}, {Shen},
  {Wu}, {Anderson}, {Schneider}, {Vanden Berk}, {Gallagher}, {Fan}, \&
  {York}}]{gibson09}
{Gibson}, R.~R., {Jiang}, L., {Brandt}, W.~N., {et~al.} 2009, \apj, 692, 758

\bibitem[{{Hao} {et~al.}(2005){Hao}, {Strauss}, {Tremonti}, {Schlegel},
  {Heckman}, {Kauffmann}, {Blanton}, {Fan}, {Gunn}, {Hall}, {Ivezi{\'c}},
  {Knapp}, {Krolik}, {Lupton}, {Richards}, {Schneider}, {Strateva}, {Zakamska},
  {Brinkmann}, {Brunner}, \& {Szokoly}}]{hao05}
{Hao}, L., {Strauss}, M.~A., {Tremonti}, C.~A., {et~al.} 2005, \aj, 129, 1783

\bibitem[{{Hawkins}(2004)}]{hawkins04}
{Hawkins}, M.~R.~S. 2004, \aap, 424, 519

\bibitem[{{Heidt} \& {Nilsson}(2011)}]{heidt11}
{Heidt}, J., \& {Nilsson}, K. 2011, \aap, 529, A162

\bibitem[{{Hewett} \& {Wild}(2010)}]{hewett10}
{Hewett}, P.~C., \& {Wild}, V. 2010, \mnras, 405, 2302

\bibitem[{{Hryniewicz} {et~al.}(2010){Hryniewicz}, {Czerny}, {Niko{\l}ajuk}, \&
  {Kuraszkiewicz}}]{hryniewicz10}
{Hryniewicz}, K., {Czerny}, B., {Niko{\l}ajuk}, M., \& {Kuraszkiewicz}, J.
  2010, \mnras, 404, 2028

\bibitem[{{Just} {et~al.}(2007){Just}, {Brandt}, {Shemmer}, {Steffen},
  {Schneider}, {Chartas}, \& {Garmire}}]{just07}
{Just}, D.~W., {Brandt}, W.~N., {Shemmer}, O., {et~al.} 2007, \apj, 665, 1004

\bibitem[{{Kaspi} {et~al.}(2005){Kaspi}, {Maoz}, {Netzer}, {Peterson},
  {Vestergaard}, \& {Jannuzi}}]{kaspi05}
{Kaspi}, S., {Maoz}, D., {Netzer}, H., {et~al.} 2005, \apj, 629, 61

\bibitem[{{Kellermann} {et~al.}(1989){Kellermann}, {Sramek}, {Schmidt},
  {Shaffer}, \& {Green}}]{kellermann89}
{Kellermann}, K.~I., {Sramek}, R., {Schmidt}, M., {Shaffer}, D.~B., \& {Green},
  R. 1989, \aj, 98, 1195

\bibitem[{{Kratzer} \& {Richards}(2014)}]{kratzer14}
{Kratzer}, R.~M., \& {Richards}, G.~T. 2014, arXiv:1405.2344

\bibitem[{{Kruczek} {et~al.}(2011){Kruczek}, {Richards}, {Gallagher}, {Deo},
  {Hall}, {Hewett}, {Leighly}, {Krawczyk}, \& {Proga}}]{kruczek11}
{Kruczek}, N.~E., {Richards}, G.~T., {Gallagher}, S.~C., {et~al.} 2011, \aj,
  142, 130

\bibitem[{{Lane} {et~al.}(2011){Lane}, {Shemmer}, {Diamond-Stanic}, {Fan},
  {Anderson}, {Brandt}, {Plotkin}, {Richards}, {Schneider}, \&
  {Strauss}}]{lane11}
{Lane}, R.~A., {Shemmer}, O., {Diamond-Stanic}, A.~M., {et~al.} 2011, \apj,
  743, 163

\bibitem[{{Laor} \& {Davis}(2011)}]{laor11}
{Laor}, A., \& {Davis}, S.~W. 2011, \mnras, 417, 681

\bibitem[{{Laor} {et~al.}(1997){Laor}, {Fiore}, {Elvis}, {Wilkes}, \&
  {McDowell}}]{laor97}
{Laor}, A., {Fiore}, F., {Elvis}, M., {Wilkes}, B.~J., \& {McDowell}, J.~C.
  1997, \apj, 477, 93

\bibitem[{{Leighly}(2004)}]{leighly04}
{Leighly}, K.~M. 2004, \apj, 611, 125

\bibitem[Leighly 
\& Moore(2004)]{leighly04a} Leighly, K.~M., \& Moore, J.~R.\ 2004, \apj, 611, 107 

\bibitem[{{Leighly} {et~al.}(2007{\natexlab{a}}){Leighly}, {Halpern},
  {Jenkins}, \& {Casebeer}}]{leighly07a}
{Leighly}, K.~M., {Halpern}, J.~P., {Jenkins}, E.~B., \& {Casebeer}, D.
  2007{\natexlab{a}}, \apjs, 173, 1

\bibitem[{{Leighly} {et~al.}(2007{\natexlab{b}}){Leighly}, {Halpern},
  {Jenkins}, {Grupe}, {Choi}, \& {Prescott}}]{leighly07}
{Leighly}, K.~M., {Halpern}, J.~P., {Jenkins}, E.~B., {et~al.}
  2007{\natexlab{b}}, \apj, 663, 103

\bibitem[{{Liu} \& {Zhang}(2011)}]{liu11}
{Liu}, Y., \& {Zhang}, S.~N. 2011, \apjl, 728, L44

\bibitem[{{Luo} {et~al.}(2015)}]{luo15}
{Luo}, B., {et~al.} 2015, ApJ, subm.

\bibitem[{{Marconi} {et~al.}(2004){Marconi}, {Risaliti}, {Gilli}, {Hunt},
  {Maiolino}, \& {Salvati}}]{marconi04}
{Marconi}, A., {Risaliti}, G., {Gilli}, R., {et~al.} 2004, \mnras, 351, 169

\bibitem[{{Markwardt}(2009)}]{markwardt09}
{Markwardt}, C.~B. 2009, in Astronomical Society of the Pacific Conference
  Series, Vol. 411, Astronomical Data Analysis Software and Systems XVIII, ed.
  D.~A. {Bohlender}, D.~{Durand}, \& P.~{Dowler}, 251

\bibitem[{{McDowell} {et~al.}(1995){McDowell}, {Canizares}, {Elvis},
  {Lawrence}, {Markoff}, {Mathur}, \& {Wilkes}}]{mcdowell95}
{McDowell}, J.~C., {Canizares}, C., {Elvis}, M., {et~al.} 1995, \apj, 450, 585

\bibitem[{{Meusinger} \& {Balafkan}(2014)}]{meusinger14}
{Meusinger}, H., \& {Balafkan}, N. 2014, \aap, 568, A114

\bibitem[{{Modigliani} {et~al.}(2010){Modigliani}, {Goldoni}, {Royer},
  {Haigron}, {Guglielmi}, {Fran{\c c}ois}, {Horrobin}, {Bristow}, {Vernet},
  {Moehler}, {Kerber}, {Ballester}, {Mason}, \& {Christensen}}]{modigliani10}
{Modigliani}, A., {Goldoni}, P., {Royer}, F., {et~al.} 2010, in Society of
  Photo-Optical Instrumentation Engineers (SPIE) Conference Series, Vol. 7737,
  Society of Photo-Optical Instrumentation Engineers (SPIE) Conference Series,
  28

\bibitem[{{Morokuma} {et~al.}(2007){Morokuma}, {Inada}, {Oguri}, {Ichikawa},
  {Kawano}, {Tokita}, {Kayo}, {Hall}, {Kochanek}, {Richards}, {York}, \&
  {Schneider}}]{morokuma07}
{Morokuma}, T., {Inada}, N., {Oguri}, M., {et~al.} 2007, \aj, 133, 214

\bibitem[{{Murray} \& {Chiang}(1995)}]{murray95a}
{Murray}, N., \& {Chiang}, J. 1995, \apjl, 454, L105

\bibitem[{{Murray} {et~al.}(1995){Murray}, {Chiang}, {Grossman}, \&
  {Voit}}]{murray95}
{Murray}, N., {Chiang}, J., {Grossman}, S.~A., \& {Voit}, G.~M. 1995, \apj,
  451, 498

\bibitem[{{Netzer} {et~al.}(2007){Netzer}, {Lira}, {Trakhtenbrot}, {Shemmer},
  \& {Cury}}]{netzer07a}
{Netzer}, H., {Lira}, P., {Trakhtenbrot}, B., {Shemmer}, O., \& {Cury}, I.
  2007, \apj, 671, 1256

\bibitem[{{Netzer} {et~al.}(2004){Netzer}, {Shemmer}, {Maiolino}, {Oliva},
  {Croom}, {Corbett}, \& {di Fabrizio}}]{netzer04}
{Netzer}, H., {Shemmer}, O., {Maiolino}, R., {et~al.} 2004, \apj, 614, 558

\bibitem[{{Netzer} \& {Trakhtenbrot}(2014)}]{netzer14}
{Netzer}, H., \& {Trakhtenbrot}, B. 2014, \mnras, 438, 672

\bibitem[{{Nicastro}(2000)}]{nicastro00}
{Nicastro}, F. 2000, \apjl, 530, L65

\bibitem[{{Nicastro} {et~al.}(2003){Nicastro}, {Martocchia}, \&
  {Matt}}]{nicastro03}
{Nicastro}, F., {Martocchia}, A., \& {Matt}, G. 2003, \apjl, 589, L13

\bibitem[{{Niko{\l}ajuk} \& {Walter}(2012)}]{nikoajuk12}
{Niko{\l}ajuk}, M., \& {Walter}, R. 2012, \mnras, 420, 2518

\bibitem[{{Peterson}(1993)}]{peterson93}
{Peterson}, B.~M. 1993, \pasp, 105, 247

\bibitem[{{Pita} {et~al.}(2014){Pita}, {Goldoni}, {Boisson}, {Lenain}, {Punch},
  {G{\'e}rard}, {Hammer}, {Kaper}, \& {Sol}}]{pita14}
{Pita}, S., {Goldoni}, P., {Boisson}, C., {et~al.} 2014, \aap, 565, A12


\bibitem[{{Plotkin} {et~al.}(2010{\natexlab{a}}){Plotkin}, {Anderson},
  {Brandt}, {Diamond-Stanic}, {Fan}, {Hall}, {Kimball}, {Richmond},
  {Schneider}, {Shemmer}, {Voges}, {York}, {Bahcall}, {Snedden}, {Bizyaev},
  {Brewington}, {Malanushenko}, {Malanushenko}, {Oravetz}, {Pan}, \&
  {Simmons}}]{plotkin10}
---. 2010{\natexlab{a}}, \aj, 139, 390 (P10)

\bibitem[{{Plotkin} {et~al.}(2010{\natexlab{b}}){Plotkin}, {Anderson},
  {Brandt}, {Diamond-Stanic}, {Fan}, {MacLeod}, {Schneider}, \&
  {Shemmer}}]{plotkin10a}
{Plotkin}, R.~M., {Anderson}, S.~F., {Brandt}, W.~N., {et~al.}
  2010{\natexlab{b}}, \apj, 721, 562

\bibitem[{{Plotkin} {et~al.}(2012){Plotkin}, {Anderson}, {Brandt}, {Markoff},
  {Shemmer}, \& {Wu}}]{plotkin12}
---. 2012, \apjl, 745, L27


\bibitem[{{Richards} {et~al.}(2002){Richards}, {Fan}, {Newberg}, {Strauss},
  {Vanden Berk}, {Schneider}, {Yanny}, {Boucher}, {Burles}, {Frieman}, {Gunn},
  {Hall}, {Ivezi{\'c}}, {Kent}, {Loveday}, {Lupton}, {Rockosi}, {Schlegel},
  {Stoughton}, {SubbaRao}, \& {York}}]{richards02}
{Richards}, G.~T., {Fan}, X., {Newberg}, H.~J., {et~al.} 2002, \aj, 123, 2945

\bibitem[{{Richards} {et~al.}(2011){Richards}, {Kruczek}, {Gallagher}, {Hall},
  {Hewett}, {Leighly}, {Deo}, {Kratzer}, \& {Shen}}]{richards11}
{Richards}, G.~T., {Kruczek}, N.~E., {Gallagher}, S.~C., {et~al.} 2011, \aj,
  141, 167

\bibitem[{{Ruan} {et~al.}(2014){Ruan}, {Anderson}, {Plotkin}, {Brandt},
  {Burnett}, {Myers}, \& {Schneider}}]{ruan14}
{Ruan}, J.~J., {Anderson}, S.~F., {Plotkin}, R.~M., {et~al.} 2014, \apj, 797,
  19

\bibitem[{{Schlegel} {et~al.}(1998){Schlegel}, {Finkbeiner}, \&
  {Davis}}]{schlegel98}
{Schlegel}, D.~J., {Finkbeiner}, D.~P., \& {Davis}, M. 1998, \apj, 500, 525

\bibitem[{{Schneider} {et~al.}(2010){Schneider}, {Richards}, {Hall}, {Strauss},
  {Anderson}, {Boroson}, {Ross}, {Shen}, {Brandt}, {Fan}, {Inada}, {Jester},
  {Knapp}, {Krawczyk}, {Thakar}, {Vanden Berk}, {Voges}, {Yanny}, {York},
  {Bahcall}, {Bizyaev}, {Blanton}, {Brewington}, {Brinkmann}, {Eisenstein},
  {Frieman}, {Fukugita}, {Gray}, {Gunn}, {Hibon}, {Ivezi{\'c}}, {Kent}, {Kron},
  {Lee}, {Lupton}, {Malanushenko}, {Malanushenko}, {Oravetz}, {Pan}, {Pier},
  {Price}, {Saxe}, {Schlegel}, {Simmons}, {Snedden}, {SubbaRao}, {Szalay}, \&
  {Weinberg}}]{schneider10}
{Schneider}, D.~P., {Richards}, G.~T., {Hall}, P.~B., {et~al.} 2010, \aj, 139,
  2360

\bibitem[{{Shang} {et~al.}(2007){Shang}, {Wills}, {Wills}, \&
  {Brotherton}}]{shang07}
{Shang}, Z., {Wills}, B.~J., {Wills}, D., \& {Brotherton}, M.~S. 2007, \aj,
  134, 294

\bibitem[{{Shang} {et~al.}(2011){Shang}, {Brotherton}, {Wills}, {Wills},
  {Cales}, {Dale}, {Green}, {Runnoe}, {Nemmen}, {Gallagher}, {Ganguly},
  {Hines}, {Kelly}, {Kriss}, {Li}, {Tang}, \& {Xie}}]{shang11}
{Shang}, Z., {Brotherton}, M.~S., {Wills}, B.~J., {et~al.} 2011, \apjs, 196, 2

\bibitem[{{Shemmer} {et~al.}(2009){Shemmer}, {Brandt}, {Anderson},
  {Diamond-Stanic}, {Fan}, {Richards}, {Schneider}, \& {Strauss}}]{shemmer09}
{Shemmer}, O., {Brandt}, W.~N., {Anderson}, S.~F., {et~al.} 2009, \apj, 696,
  580

\bibitem[{{Shemmer} \& {Lieber}(2015)}]{shemmer15}
{Shemmer}, O., \& {Lieber}, S. 2015, ApJ, subm.

\bibitem[{{Shemmer} {et~al.}(2004){Shemmer}, {Netzer}, {Maiolino}, {Oliva},
  {Croom}, {Corbett}, \& {di Fabrizio}}]{shemmer04}
{Shemmer}, O., {Netzer}, H., {Maiolino}, R., {et~al.} 2004, \apj, 614, 547

\bibitem[{{Shemmer} {et~al.}(2006){Shemmer}, {Brandt}, {Schneider}, {Fan},
  {Strauss}, {Diamond-Stanic}, {Richards}, {Anderson}, {Gunn}, \&
  {Brinkmann}}]{shemmer06}
{Shemmer}, O., {Brandt}, W.~N., {Schneider}, D.~P., {et~al.} 2006, \apj, 644,
  86

\bibitem[{{Shemmer} {et~al.}(2010){Shemmer}, {Trakhtenbrot}, {Anderson},
  {Brandt}, {Diamond-Stanic}, {Fan}, {Lira}, {Netzer}, {Plotkin}, {Richards},
  {Schneider}, \& {Strauss}}]{shemmer10}
{Shemmer}, O., {Trakhtenbrot}, B., {Anderson}, S.~F., {et~al.} 2010, \apjl,
  722, L152

\bibitem[{{Shen} \& {Ho}(2014)}]{shen14}
{Shen}, Y., \& {Ho}, L.~C. 2014, \nat, 513, 210

\bibitem[{{Shen} {et~al.}(2011){Shen}, {Richards}, {Strauss}, {Hall},
  {Schneider}, {Snedden}, {Bizyaev}, {Brewington}, {Malanushenko},
  {Malanushenko}, {Oravetz}, {Pan}, \& {Simmons}}]{shen11}
{Shen}, Y., {Richards}, G.~T., {Strauss}, M.~A., {et~al.} 2011, \apjs, 194, 45

\bibitem[{{Sheskin}(2011)}]{sheskin11}{Sheskin}, D.~J. 2011, Handbook of Parametric and Nonparametric Statistical Procedures, Fifth Edition

\bibitem[{{Smith} {et~al.}(2007){Smith}, {Williams}, {Schmidt},
  {Diamond-Stanic}, \& {Means}}]{smith07}
{Smith}, P.~S., {Williams}, G.~G., {Schmidt}, G.~D., {Diamond-Stanic}, A.~M.,
  \& {Means}, D.~L. 2007, \apj, 663, 118

\bibitem[{{Sulentic} {et~al.}(2007){Sulentic}, {Bachev}, {Marziani}, {Negrete},
  \& {Dultzin}}]{sulentic07}
{Sulentic}, J.~W., {Bachev}, R., {Marziani}, P., {Negrete}, C.~A., \&
  {Dultzin}, D. 2007, \apj, 666, 757

\bibitem[{{Sulentic} {et~al.}(2000){Sulentic}, {Zwitter}, {Marziani}, \&
  {Dultzin-Hacyan}}]{sulentic00}
{Sulentic}, J.~W., {Zwitter}, T., {Marziani}, P., \& {Dultzin-Hacyan}, D. 2000,
  \apjl, 536, L5

\bibitem[{{Tang} {et~al.}(2012){Tang}, {Shang}, {Gu}, {Brotherton}, \&
  {Runnoe}}]{tang12}
{Tang}, B., {Shang}, Z., {Gu}, Q., {Brotherton}, M.~S., \& {Runnoe}, J.~C.
  2012, \apjs, 201, 38

\bibitem[{{Trakhtenbrot}(2014)}]{trakhtenbrot14}
{Trakhtenbrot}, B. 2014, \apjl, 789, L9

\bibitem[{{Trakhtenbrot} \& {Netzer}(2012)}]{trakhtenbrot12}
{Trakhtenbrot}, B., \& {Netzer}, H. 2012, \mnras, 427, 3081

\bibitem[{{Tran} {et~al.}(2011){Tran}, {Lyke}, \& {Mader}}]{tran11}
{Tran}, H.~D., {Lyke}, J.~E., \& {Mader}, J.~A. 2011, \apjl, 726, L21

\bibitem[{{Trump} {et~al.}(2011){Trump}, {Impey}, {Kelly}, {Civano}, {Gabor},
  {Diamond-Stanic}, {Merloni}, {Urry}, {Hao}, {Jahnke}, {Nagao}, {Taniguchi},
  {Koekemoer}, {Lanzuisi}, {Liu}, {Mainieri}, {Salvato}, \&
  {Scoville}}]{trump11}
{Trump}, J.~R., {Impey}, C.~D., {Kelly}, B.~C., {et~al.} 2011, \apj, 733, 60

\bibitem[{{Vacca} {et~al.}(2003){Vacca}, {Cushing}, \& {Rayner}}]{vacca03}
{Vacca}, W.~D., {Cushing}, M.~C., \& {Rayner}, J.~T. 2003, \pasp, 115, 389

\bibitem[{{van Dokkum}(2001)}]{van-dokkum01}
{van Dokkum}, P.~G. 2001, \pasp, 113, 1420

\bibitem[{{Vanden Berk} {et~al.}(2001){Vanden Berk}, {Richards}, {Bauer},
  {Strauss}, {Schneider}, {Heckman}, {York}, {Hall}, {Fan}, {Knapp},
  {Anderson}, {Annis}, {Bahcall}, {Bernardi}, {Briggs}, {Brinkmann}, {Brunner},
  {Burles}, {Carey}, {Castander}, {Connolly}, {Crocker}, {Csabai}, {Doi},
  {Finkbeiner}, {Friedman}, {Frieman}, {Fukugita}, {Gunn}, {Hennessy},
  {Ivezi{\'c}}, {Kent}, {Kunszt}, {Lamb}, {Leger}, {Long}, {Loveday}, {Lupton},
  {Meiksin}, {Merelli}, {Munn}, {Newberg}, {Newcomb}, {Nichol}, {Owen}, {Pier},
  {Pope}, {Rockosi}, {Schlegel}, {Siegmund}, {Smee}, {Snir}, {Stoughton},
  {Stubbs}, {SubbaRao}, {Szalay}, {Szokoly}, {Tremonti}, {Uomoto}, {Waddell},
  {Yanny}, \& {Zheng}}]{vanden-berk01}
{Vanden Berk}, D.~E., {Richards}, G.~T., {Bauer}, A., {et~al.} 2001, \aj, 122,
  549

\bibitem[{{Vernet} {et~al.}(2011){Vernet}, {Dekker}, {D'Odorico}, {Kaper},
  {Kjaergaard}, {Hammer}, {Randich}, {Zerbi}, {Groot}, {Hjorth}, {Guinouard},
  {Navarro}, {Adolfse}, {Albers}, {Amans}, {Andersen}, {Andersen}, {Binetruy},
  {Bristow}, {Castillo}, {Chemla}, {Christensen}, {Conconi}, {Conzelmann},
  {Dam}, {de Caprio}, {de Ugarte Postigo}, {Delabre}, {di Marcantonio},
  {Downing}, {Elswijk}, {Finger}, {Fischer}, {Flores}, {Fran{\c c}ois},
  {Goldoni}, {Guglielmi}, {Haigron}, {Hanenburg}, {Hendriks}, {Horrobin},
  {Horville}, {Jessen}, {Kerber}, {Kern}, {Kiekebusch}, {Kleszcz}, {Klougart},
  {Kragt}, {Larsen}, {Lizon}, {Lucuix}, {Mainieri}, {Manuputy}, {Martayan},
  {Mason}, {Mazzoleni}, {Michaelsen}, {Modigliani}, {Moehler}, {M{\o}ller},
  {Norup S{\o}rensen}, {N{\o}rregaard}, {P{\'e}roux}, {Patat}, {Pena}, {Pragt},
  {Reinero}, {Rigal}, {Riva}, {Roelfsema}, {Royer}, {Sacco}, {Santin},
  {Schoenmaker}, {Spano}, {Sweers}, {Ter Horst}, {Tintori}, {Tromp}, {van
  Dael}, {van der Vliet}, {Venema}, {Vidali}, {Vinther}, {Vola}, {Winters},
  {Wistisen}, {Wulterkens}, \& {Zacchei}}]{vernet11}
{Vernet}, J., {Dekker}, H., {D'Odorico}, S., {et~al.} 2011, \aap, 536, A105

\bibitem[{{Vestergaard} \& {Wilkes}(2001)}]{vestergaard01}
{Vestergaard}, M., \& {Wilkes}, B.~J. 2001, \apjs, 134, 1

\bibitem[{{Wills} {et~al.}(1999){Wills}, {Laor}, {Brotherton}, {Wills},
  {Wilkes}, {Ferland}, \& {Shang}}]{wills99}
{Wills}, B.~J., {Laor}, A., {Brotherton}, M.~S., {et~al.} 1999, \apjl, 515, L53

\bibitem[{{Wu} {et~al.}(2012){Wu}, {Brandt}, {Anderson}, {Diamond-Stanic},
  {Hall}, {Plotkin}, {Schneider}, \& {Shemmer}}]{wu12}
{Wu}, J., {Brandt}, W.~N., {Anderson}, S.~F., {et~al.} 2012, \apj, 747, 10

\bibitem[{{Wu} {et~al.}(2011){Wu}, {Brandt}, {Hall}, {Gibson}, {Richards},
  {Schneider}, {Shemmer}, {Just}, \& {Schmidt}}]{wu11}
{Wu}, J., {Brandt}, W.~N., {Hall}, P.~B., {et~al.} 2011, \apj, 736, 28

\bibitem[{{York} {et~al.}(2000){York}, {Adelman}, {Anderson}, {Anderson},
  {Annis}, {Bahcall}, {Bakken}, {Barkhouser}, {Bastian}, {Berman}, {Boroski},
  {Bracker}, {Briegel}, {Briggs}, {Brinkmann}, {Brunner}, {Burles}, {Carey},
  {Carr}, {Castander}, {Chen}, {Colestock}, {Connolly}, {Crocker}, {Csabai},
  {Czarapata}, {Davis}, {Doi}, {Dombeck}, {Eisenstein}, {Ellman}, {Elms},
  {Evans}, {Fan}, {Federwitz}, {Fiscelli}, {Friedman}, {Frieman}, {Fukugita},
  {Gillespie}, {Gunn}, {Gurbani}, {de Haas}, {Haldeman}, {Harris}, {Hayes},
  {Heckman}, {Hennessy}, {Hindsley}, {Holm}, {Holmgren}, {Huang}, {Hull},
  {Husby}, {Ichikawa}, {Ichikawa}, {Ivezi{\'c}}, {Kent}, {Kim}, {Kinney},
  {Klaene}, {Kleinman}, {Kleinman}, {Knapp}, {Korienek}, {Kron}, {Kunszt},
  {Lamb}, {Lee}, {Leger}, {Limmongkol}, {Lindenmeyer}, {Long}, {Loomis},
  {Loveday}, {Lucinio}, {Lupton}, {MacKinnon}, {Mannery}, {Mantsch}, {Margon},
  {McGehee}, {McKay}, {Meiksin}, {Merelli}, {Monet}, {Munn}, {Narayanan},
  {Nash}, {Neilsen}, {Neswold}, {Newberg}, {Nichol}, {Nicinski}, {Nonino},
  {Okada}, {Okamura}, {Ostriker}, {Owen}, {Pauls}, {Peoples}, {Peterson},
  {Petravick}, {Pier}, {Pope}, {Pordes}, {Prosapio}, {Rechenmacher}, {Quinn},
  {Richards}, {Richmond}, {Rivetta}, {Rockosi}, {Ruthmansdorfer}, {Sandford},
  {Schlegel}, {Schneider}, {Sekiguchi}, {Sergey}, {Shimasaku}, {Siegmund},
  {Smee}, {Smith}, {Snedden}, {Stone}, {Stoughton}, {Strauss}, {Stubbs},
  {SubbaRao}, {Szalay}, {Szapudi}, {Szokoly}, {Thakar}, {Tremonti}, {Tucker},
  {Uomoto}, {Vanden Berk}, {Vogeley}, {Waddell}, {Wang}, {Watanabe},
  {Weinberg}, {Yanny}, {Yasuda}, \& {SDSS Collaboration}}]{york00}
{York}, D.~G., {Adelman}, J., {Anderson}, Jr., J.~E., {et~al.} 2000, \aj, 120,
  1579

\end{thebibliography}

\newpage

\tablecaption{X-shooter Observation Log}
\tabletypesize{\footnotesize}
\tablecolumns{8}
\tablewidth{0pt}
\renewcommand{\arraystretch}{1.2}
\begin{deluxetable*}{l c c c c c l c c c c}
\tablehead{
	\colhead{Source Name} & 
	\colhead{} & 
	\colhead{} & 
	\colhead{} & 
	\colhead{$i_{\rm psf}$\tablenotemark{d}} & 
	\colhead{$M_i$\tablenotemark{e}} & 
	\colhead{} & 
	\multicolumn{3}{c}{Exp. Time (min)} \\ 
	\colhead{(SDSS J)} & 
	\colhead{$z$\tablenotemark{a}} & 
	\colhead{$z_{\rm sys}$\tablenotemark{b}} & 
	\colhead{$\alpha_{\lambda}$\tablenotemark{c}} & 
	\colhead{(mag)} & 
	\colhead{(mag)} & 
	\colhead{Obs. Date} &
	\colhead{UVB} & 
	\colhead{VIS}  & 
	\colhead{NIR} \\ 
}
\startdata
083650.86$+$142539.0 &
1.750   &  1.749  & $-1.1$ & 18.72 & $-26.69$ & 2012 Feb 25  & 38 & 40 & 40  \\

094533.98$+$100950.1 &
1.671   &  1.683  & $-1.1$ & 17.44 & $-27.80$ & 2011 Dec 21  & 20 & 20 & 20 \\

132138.86$+$010846.3 &
1.423   &  1.422 & $-1.7$ & 18.80 & $-26.08$ & 2013 Mar 4 & 38 & 40 & 40 \\

133222.62$+$034739.9\tablenotemark{f} &
1.447   &  1.442  & $-0.7$ & 17.94 & $-26.97$ & 2013 Mar 3  & 20 & 22 & 20 \\

141141.96$+$140233.9 &
1.753   &  1.754  & $-1.8$ & 18.53 &$-26.84$ & 2013 Feb 12  & 38 & 40 & 40 \\

141730.92$+$073320.7 &
1.710   &  1.716  & $-1.7$ & 18.40 & $-26.92$ & 2013 Mar 4  & 38 & 40 & 40 \\

144741.76$-$020339.1 &
1.431   &  1.430  & $-2.3$ & 18.61 & $-26.37$ & 2012 Mar 23  & 38 & 40 & 40 \\
\enddata
\tablenotetext{a}{Redshift obtained from Hewett \& Wild (2010).}
\tablenotetext{b}{Systemic redshift (see Section~\ref{sec:redshift}).}
\tablenotetext{c}{Best-fit power-law index ($f_{\lambda} \sim \lambda^{\alpha_\lambda}$) for the continuum of each X-shooter spectrum, fit between rest-frame 1680-1710, 1975-2050, 2150-2250, and 4010-4050~\AA\ (see Section~\ref{sec:pl}).}
\tablenotetext{d}{SDSS PSF $i$-band magnitude from \citet{schneider10}.}
\tablenotetext{e}{Absolute $i$-band magnitude, K-corrected to $z=2$, from \citet{shen11} (using the same cosmology as adopted here).  These absolute magnitudes include both continuum and emission line K-corrections, for which the latter may be incorrect for WLQs.  We adopt these $M_i$ values nonetheless, for consistency with comparison quasar samples for which we also utilize measurements from \citet{shen11}.}
\tablenotetext{f}{A known lensed quasar; see Morokuma \et (2007) and Section~\ref{sec:gravlens}.}
\label{tab:obslog}
\end{deluxetable*}
\renewcommand{\arraystretch}{1.}



\tablecaption{Rest-frame Equivalent Width Measures}
\tabletypesize{\footnotesize}
\tablecolumns{9}
\tablewidth{0pt}
\renewcommand{\arraystretch}{1.2}
\begin{deluxetable*}{l l l l l l l l l}
\tablehead{
	\colhead{Name} &  
	\colhead{H$\alpha$}        & 
	\colhead{H$\beta$}        & 
	\colhead{\ion{Mg}{2}}        & 
	\colhead{\ion{C}{4}}          & 
	\colhead{\ion{Fe}{2}\tablenotemark{a}}  & 
	\colhead{\ion{Fe}{2}[UV]\tablenotemark{b}} & 
	\colhead{\ion{He}{2}$\lambda$4686\tablenotemark{c}}  & 
	\colhead{\ion{He}{2}$\lambda$1640\tablenotemark{d}} \\%
	\colhead{}  & 
	\colhead{(\AA)}          & 
	\colhead{(\AA)}          & 
	\colhead{(\AA)}          & 
	\colhead{(\AA)}          & 
	\colhead{(\AA)}          & 
	\colhead{(\AA)}          & 
	\colhead{(\AA)}          & 
	\colhead{(\AA)}     \\
}
\startdata
       J0836 &        $162.2^{+40.6}_{-13.7}$ &          $27.9^{+51.3}_{-4.5}$ &            $9.6^{+1.4}_{-0.7}$ &            $4.2^{+0.3}_{-0.5}$ &                $69.2 \pm 16.9$ &                 $14.2 \pm 5.9$ &  $<$8.0   &  $<$0.4\\
     J0945 &        $210.6^{+39.4}_{-18.5}$ &           $31.8^{+5.1}_{-3.6}$ &           $18.3^{+0.3}_{-0.5}$ &            $2.9^{+0.3}_{-0.6}$ &                 $63.6 \pm 9.5$ &                 $35.7 \pm 2.6$  &   $<$1.3  &  $<$0.2  \\
     J1321 &         $236.3^{+3.5}_{-18.9}$ &          $37.5^{+17.8}_{-4.2}$ &           $16.1^{+3.8}_{-0.4}$ &           $18.7^{+0.8}_{-1.6}$ &          $32.6^{+32.1}_{-7.8}$ &          $27.6^{+15.9}_{-0.9}$  & \nodata  &  $0.9\pm0.1$ \\
     J1411 &         $183.4^{+1.2}_{-24.2}$ &                $32.3 \pm 10.2$ &            $6.3^{+0.6}_{-0.4}$ &            $3.8^{+0.8}_{-0.2}$ &                $45.5 \pm 17.0$ &                 $7.1 \pm 2.8$ & $<$7.5  &  $<$0.9 \\
     J1417 &         $151.8^{+18.4}_{-1.3}$ &                 $15.6 \pm 2.4$ &           $12.9^{+0.5}_{-0.2}$ &            $2.5^{+2.1}_{-0.7}$ &                 $25.7 \pm 4.4$ &                 $27.7 \pm 1.9$ & $<$7.0   &  $0.2\pm0.07$ \\
     J1447 &         $237.9^{+6.8}_{-18.1}$ &          $31.2^{+16.0}_{-2.6}$ &           $13.1^{+2.5}_{-0.1}$ &            $7.7^{+0.2}_{-1.3}$ &          $49.8^{+21.4}_{-5.2}$ &          $13.2^{+13.3}_{-2.7}$ & \nodata  & $0.8\pm0.06$ \\
 
 \multicolumn{7}{l}{Other WLQs  with \hb\ in the Literature} \\
 SDSS J114153.34$+$021924.3\tablenotemark{e}   &    \nodata    &    $20^{+4}_{-3}$     &   \nodata   &  $0.4\pm0.2$  &  \nodata    &   \nodata   & \nodata & \nodata \\
  SDSS J123743.08$+$630144.9\tablenotemark{e}   &    \nodata    &    $35^{+8}_{-5}$     &   \nodata   &  $7.7\pm1.1$        &   \nodata  &   \nodata & \nodata & \nodata \\
  SDSS J152156.48+520238.5\tablenotemark{f}    &    \nodata    &    30.7                         &   \nodata   &  $9.1\pm0.6$ &   \nodata                        &   \nodata  & \nodata & \nodata \\
  PG1407\tablenotemark{g} &   \nodata          &    $<$40                     & \nodata            &  $4.6\pm2$ &       \nodata               &  \nodata & \nodata & \nodata \\
PHL1811\tablenotemark{h} &  \nodata          &    50                           & \nodata         &  6.6            &       \nodata              &  \nodata  & \nodata & \nodata \\
\enddata
\tablenotetext{a}{Equivalent width for optical Fe is calculated between 4434-4684~\AA.}
\tablenotetext{b}{Equivalent width for UV Fe is calculated between 2250-2650~\AA.}
\tablenotetext{c}{\ion{He}{2}$\lambda$4686 is measured between 4665-4700~\AA.  Upper limits are quoted at the 3$\sigma$ level (see Section~\ref{sec:disc:softsed:heii})}. 
\tablenotetext{d}{\ion{He}{2}$\lambda$1640 is measured between 1620-1650~\AA.  Upper limits are quoted at the 3$\sigma$ level, and errors on detected \uvheii\ are quoted at the 1$\sigma$ level (see Section~\ref{sec:disc:softsed:heii}).}
\tablenotetext{e}{\rew(\hb) is from \citet{shemmer10}; \rew[\civ] is from \citetalias{diamond-stanic09}.  \citet{shen11} report \rew[\civ]=$5.3\pm2.0$~\AA\ for \srcshtwofull.}
\tablenotetext{f}{\rew(\hb) is from \citet{wu11}; \rew[\civ] is from \citet{wu11}.  \citet{shen11} report \rew[\civ]=$3.0\pm0.2$~\AA.}
\tablenotetext{g}{Line measurements from \citet{mcdowell95}.  They present line measurements across two epochs for \civ, and we report the epoch closest in time (1992) to their \hb\ observations (1994).  The other \civ\ epoch has \rew$=3.6\pm2.5$~\AA.}
\tablenotetext{h}{Line measurements from \citet{leighly07a}.}
\label{tab:rew}
\end{deluxetable*}
\renewcommand{\arraystretch}{1.}
\clearpage


\tablecaption{FWHM Measures}
\tabletypesize{\footnotesize}
\tablecolumns{8}
\tablewidth{0pt}
\renewcommand{\arraystretch}{1.2}
\begin{deluxetable*}{l l c l c l c l}
\tablehead{
	\colhead{Name} &  
	\colhead{\ha}        & 
	\colhead{\ion{Fe}{2}[\ha]}        & 
	\colhead{\hb}        & 
	\colhead{\ion{Fe}{2}[\hb]}        & 
	\colhead{\mgii}        & 
	\colhead{\ion{Fe}{2}[\mgii]}        & 
	\colhead{\civ}          \\ 
	\colhead{}  & 
	\colhead{(\kms)}          & 
	\colhead{(\kms)}          & 
	\colhead{(\kms)}          & 
	\colhead{(\kms)}          & 
	\colhead{(\kms)}          & 
	\colhead{(\kms)}          & 
	\colhead{(\kms)}          
}
\startdata
      J0836 &           $2298^{+287}_{-412}$ &                           1575 &         $2880^{+1877}_{-1069}$ &                           2775 &           $3284^{+695}_{-210}$ &                           3850 &           $5423^{+570}_{-853}$ \\
     J0945 &            $3368^{+112}_{-50}$ &                           3525 &                 $4278 \pm 598$ &                           1800 &            $5695^{+78}_{-199}$ &                           2950 &         $8018^{+1058}_{-1730}$ \\
     J1321 &            $2461^{+92}_{-296}$ &                        \nodata &          $2551^{+1173}_{-435}$ &                           2025 &           $3460^{+510}_{-51}$ &                           1925 &           $2760^{+488}_{-264}$ \\
     J1411 &           $3137^{+282}_{-473}$ &                        \nodata &                $3966 \pm 1256$ &                           1475 &           $2860^{+420}_{-170}$ &                           1075 &          $6177^{+2210}_{-410}$ \\
     J1417 &            $1973^{+140}_{-31}$ &                           2250 &                 $2784 \pm 759$ &                            925 &            $4717^{+222}_{-51}$ &                           1375 &        $9694^{+10305}_{-2980}$ \\
     J1447 &            $2000^{+19}_{-209}$ &                        \nodata &           $1923^{+933}_{-164}$ &                           1500 &            $2442^{+618}_{-32}$ &                           1975 &          $4536^{+1539}_{-991}$ \\
      \enddata

\label{tab:fwhm}
\end{deluxetable*}
\renewcommand{\arraystretch}{1.}


\tablecaption{Line Luminosities and Blueshifts}
\tabletypesize{\footnotesize}
\tablecolumns{7}
\tablewidth{0pt}
\renewcommand{\arraystretch}{1.2}
\begin{deluxetable*}{l l l l l r r}
\tablehead{
	\colhead{Name} &  
	\colhead{log L[\ha]}        & 
	\colhead{log L[\hb]}        & 
	\colhead{log L[\mgii]}        & 
	\colhead{log L[\civ]}        & 
	\colhead{$\Delta v$[\mgii]}  & 
	\colhead{$\Delta v$[\civ]}          \\ 
	\colhead{}  & 
	\colhead{(\ergs)}          & 
	\colhead{(\ergs)}          & 
	\colhead{(\ergs)}          & 
	\colhead{(\ergs)}          & 
	\colhead{(\kms)}          & 
	\colhead{(\kms)}          
}
\startdata
      J0836 &        $44.29^{+0.21}_{-0.06}$ &        $43.71^{+1.09}_{-0.19}$ &        $43.56^{+0.09}_{-0.06}$ &               $43.54 \pm 0.07$ &                   $197 \pm 216$ &                $2266 \pm 191$ \\
     J0945 &               $44.64 \pm 0.05$ &               $44.02 \pm 0.08$ &               $44.29 \pm 0.02$ &               $43.75 \pm 0.12$ &                $1281 \pm 183$ &                $5485 \pm 380$ \\
     J1321 &        $43.89^{+0.03}_{-0.04}$ &        $43.32^{+0.39}_{-0.09}$ &        $43.48^{+0.11}_{-0.02}$ &        $43.95^{+0.26}_{-0.06}$ &                 $202 \pm 188$ &                 $396 \pm 189$ \\
     J1411 &        $44.03^{+0.03}_{-0.06}$ &               $43.52 \pm 0.20$ &               $43.34 \pm 0.06$ &        $43.60^{+0.16}_{-0.06}$ &                   $-136 \pm 181$ &          $3142_{-208}^{+370}$ \\
     J1417 &        $44.15^{+0.05}_{-0.03}$ &               $43.45 \pm 0.11$ &               $43.88 \pm 0.02$ &        $43.55^{+2.38}_{-0.23}$ &                 $624 \pm 180$ &         $5321_{-872}^{+4178}$ \\
     J1447 &        $43.99^{+0.02}_{-0.03}$ &        $43.41^{+0.51}_{-0.07}$ &        $43.56^{+0.09}_{-0.02}$ &        $43.89^{+0.50}_{-0.15}$ &                   $1 \pm 185$ &          $1319_{-381}^{+759}$ \\

\enddata
\tablecomments{$\Delta v$[\mgii] and $\Delta v$[\civ] are line of sight blueshifts of the peaks of the \mgii\ and \civ\ profiles, respectively.  $\Delta v$ is based on the observed wavelength  of each line center (expected to be at rest-frame 2800~\AA\ for \mgii\ and 1549~\AA\ for \civ) compared to the systemic redshifts in Table~\ref{tab:obslog}.  Blueshifts are defined to be positive for approaching motions. }
\label{tab:linelum}
\end{deluxetable*}
\renewcommand{\arraystretch}{1.}


\tablecaption{Virial Black Hole Masses and Eddington Ratios}
\tabletypesize{\small}
\tablecolumns{7}
\tablewidth{0pt}
\renewcommand{\arraystretch}{1.2}
\begin{deluxetable}{l l l l l}
\tablehead{
	\colhead{Name} &  
	\colhead{$f(L)$\tablenotemark{a}}        & 
	\colhead{$\log L_{5100}$\tablenotemark{b}}        & 
	\colhead{log~\mbh \tablenotemark{c}}        & 
	\colhead{\lledd \tablenotemark{c}}        \\ 
	\colhead{}  & 
	\colhead{}          & 
	\colhead{(\ergs)}          & 
	\colhead{(\msun)}          & 
	\colhead{}         \\ 
}
\startdata
     J0836 &                           5.96 &        $45.93^{+0.10}_{-0.23}$ &         $8.59^{+0.64}_{-0.26}$ &         $0.87^{+1.36}_{-0.65}$ \\
     J0945 &                           5.85 &               $46.17 \pm 0.03$ &         $9.05^{+0.14}_{-0.08}$ &                $0.51 \pm 0.15$ \\
     J1321 &                           6.27 &        $45.41^{+0.03}_{-0.06}$ &         $8.22^{+0.47}_{-0.13}$ &         $0.63^{+0.23}_{-0.37}$ \\
     J1411 &                           6.11 &               $45.64 \pm 0.10$ &                $8.72 \pm 0.24$ &         $0.34^{+0.42}_{-0.17}$ \\
     J1417 &                           5.97 &               $45.91 \pm 0.03$ &         $8.55^{+0.30}_{-0.17}$ &                $0.92 \pm 0.50$ \\
     J1447 &                           6.16 &        $45.56^{+0.02}_{-0.06}$ &         $8.05^{+0.51}_{-0.06}$ &         $1.30^{+0.17}_{-0.78}$ \\
\enddata

\tablenotetext{a}{Bolometric correction from \citet{marconi04}.}
\tablenotetext{b}{Luminosity of the power-law continuum at 5100~\AA.  The error bars are determined in the same manner as in Section~\ref{sec:fits:errors}.}
\tablenotetext{c}{Quoted errors on log~\mbh\ and \lledd\ are based on propagating the errors determined by our spectral fitting (see \ref{sec:fits:errors}), and they do not include systematics or other uncertainties related specifically to the virial-based scaling technique (i.e., Equations~\ref{eq:mbh} and \ref{eq:lledd}).}

\label{tab:bhmass}
\end{deluxetable}
\renewcommand{\arraystretch}{1.}

\clearpage

\appendix
\section{Spectral Reduction}
Here, we describe details of the X-shooter spectral reduction.  We processed the raw data using the  X-shooter pipeline \citep{modigliani10}, with {\tt esorex} v3.10 and the X-shooter release kit v2.2.0.  Each X-shooter arm is reduced individually.  First, we create master bias (UVB/VIS) frames, as well as bad pixel maps for each arm.  Single pinhole arc calibration frames are used to make a first guess for the wavelength solution and the centers of each order.  The order centers are then traced using a single pinhole frame illuminated by a continuum lamp.  Next, master flat fields are created by combining individual frames illuminated by the continuum lamp through the slit.      A multi-pinhole arc frame is then used, along with the above first guess spectral model, to create a 2D map to determine the wavelength and spatial scale calibration of the detector.     For the UVB arm, we also calculate the instrumental response by comparing on-sky observations of the spectrophotometric star to a flux table provided by the observatory.\footnote{\url{http://www.eso.org/sci/facilities/paranal/instruments/xshooter/tools/specphot\_list.html}}

The above calibration is  applied to the individual science target frames, and to the corresponding telluric standard frames (for the VIS and NIR arms).  Each individual frame is bias/dark subtracted (when relevent) and flatfielded, bad pixels are removed, and cosmic rays are rejected (following the rejection method described by \citealt{van-dokkum01}).  The nodded frames are combined (using a ``double-pass sky subtraction'' method; see \citealt{vernet11}) to produce calibrated, background subtracted, wavelength rectified 2D spectra.  Finally, 1D spectra are extracted using a rectangular aperture centered on the source in the 2D spectra.  The size of the extraction aperture is interactively adjusted to optimize the reduction  for each source and arm.   The UVB spectra are flux calibrated, using the instrumental response calculated from  the  spectrophotometric standard star, and a slit loss correction is applied based on the seeing (typically from 0.9-1.2$\arcsec$ at 5000~\AA) and airmass (typically $\sec z \sim 1.2-1.4$) of each observation.

For the VIS/NIR arms, we  apply the telluric correction and flux calibration using the IDL routine {\tt xtellcor\_general} from the Spextool package \citep{vacca03}.  A telluric absorption spectrum is created by comparing the telluric standard's (merged 1D) spectrum to a high-resolution template for Vega (which is calibrated to the magnitude of the observed telluric star from SIMBAD and convolved to the X-shooter instrumental resolution).    Flux-calibrated, telluric-corrected spectra are created by dividing the above model by the observed telluric standard spectrum, and then multiplying by the spectrum of the science target.  Since the telluric standards were observed with the same instrument configuration as the science exposures, no additional slitloss correction is applied.     Following flux calibration,  all UVB, VIS, and NIR spectra are  dereddened for Galactic extinction, using the \citet{schlegel98} maps (typical $A_r\sim0.06-0.16$~mag) and assuming a \citet{cardelli89} reddening curve with $R_V=3.1$.

\section{Spectral Line Fitting}
All line fitting is performed in the rest-frame.  Each line complex is fit over a spectral window that is several hundred Angstroms wide.  The width of the spectral window is customized for each source/complex (see Table~\ref{tab:infit}).\footnote{For the \ha\ and \hb\ complexes, the fitting window depends on the redshift of each source (and the NIR atmospheric transmission windows); the fitting window for the \mgii\ complex is largely determined by the need to obtain enough dynamic range to be able to adequately separate the Fe emission from the local linear continuum.} 
   The local linear continuum is normalized within two 20~\AA\ bands centered on a blue wavelength ($\lambda_{\rm blue}$) and a red wavelength ($\lambda_{\rm red}$), where  Fe emission is expected to be minimal.  For the Fe continuum, we create a grid of templates convolved by a single Gaussian kernel spanning $925< {\rm FWHM} < 20000$~\kms, with the FWHM incremented in steps of 25~\kms.  The best-fitting continuum, line profile, and Fe template are chosen by stepping through each Fe template and  applying a  $\chi^2$ minimization routine.\footnote{Specifically, we use {\tt mpfit} in {\tt IDL} \citep{markwardt09}, which uses a Levenberg-Marquardt technique for (weighted) non-linear least squares fitting.}  
 In Table~\ref{tab:infit}, we summarize for each source and complex the adopted values for the spectral fitting windows, $\lambda_{\rm blue}$, $\lambda_{\rm red}$, and whether  an Fe template is required for each fit.    Four sources (\srcone, \srctwo, \srcfive, \srcsix) show narrow absorption features across the UVB (and sometimes VIS) arm.  Prior to fitting each complex, we mask these narrow absorption features by applying a sigma-clipping algorithm.  
  
Our initial fits include the above local linear continuum and  Fe continuum, to which we add up to three broad Gaussians to represent the broad emission line in each complex (i.e., \ha, \hb, \mgii, or \civ).  The individual Gaussian components are not physically meaningful.  They simply provide an analytic means to describe the line profiles, and  only 1-2 Gaussians are typically needed to model each broad emission line (see Table~\ref{tab:infit}). Each broad Gaussian is constrained to have $1200 < {\rm FWHM} < 20000$~\kms.   We then repeat each fit and include extra Gaussian components for other emission lines that are expected to fall within each fitting window. Each line complex therefore has a minimum of five free parameters, which includes two free parameters for the linear continuum (the flux density at each of the two adopted normalization wavelengths), and three free parameters for each Gaussian component (the normalization, FWHM, and central wavelength).   When an Fe continuum is included in the fit, there are two additional parameters  --- the normalization, and the FWHM of the Gaussian kernel used to broaden the Fe template.   

Other emission lines that we model include narrow components for \ha, \hb, \mgii, \civ, and additional narrow emission line region species.  For example, within the \ha\ complex, we include narrow components for  the [\ion{N}{2}] $\lambda\lambda$6548,6584 and [\ion{S}{2}] $\lambda\lambda$6717,6731 doublets, requiring the line flux ratio of the [\ion{N}{2}] doublet to be 2.96 \citep[e.g.,][]{shen11}.  Within the \hb\ complex we include narrow components for \oiii\ $\lambda\lambda$4959,5007, with the line flux ratio constrained to be 2.95. Each narrow emission line  is required to have the same FWHM (constrained to be $<1200$~\kms), and also required to have the same velocity offset from the systemic redshift.   As we add additional components to each fit, the new fit is accepted if it improves the reduced $\chi^2$ by at least 20\% \citep[see, e.g.,][]{hao05}.   We also visually examine each fit to empirically confirm if an extra component is necessary (and that the above 20\% threshold is reasonable).  Ultimately, only \srcthree\ displays firm narrow \oiii\ emission (and \srcseven\ may have tentative detections).  No other narrow emission lines are required for any other source/complex.  

\subsection{Uncertainty Estimates on Emission Line Properties}
\label{sec:fits:errors}
The uncertainties on best-fit line parameters are influenced by systematics, with the error budget dominated by uncertainties on the level of the continuum (i.e., there are degeneracies between the best-fit linear and Fe continuum levels and the parameters describing the line profiles).   Since the uncertainties are not necessarily dominated by statistical fluctuations, we  do not use the absolute value of the final reduced $\chi^2_r$ to assign  a statistical confidence to the quality of each fit, nor do we trust the error bars on each best-fit parameter returned by the $\chi^2$ minimization routine (which are likely underestimated).  We instead employ an empirical scheme to estimate uncertainties on each best-fit parameter, which implicitly accounts for  (the dominant) systematic and statistical uncertainties.  

We begin with the best spectral fit to each line complex, and we record the corresponding $\chi^2_{\rm best}$ and degrees of freedom, $\nu_{\rm best}$.  In the following, we adopt the same number of Gaussians to model the line profile as used for the above  best-fit, and we include an Fe template only if one is included in the best-fit.  We then create a $51\times51$ grid of linear continua.  Each continuum in the grid is set by a fixed flux level within the adopted 20~\AA\ continuum normalization bands,  and we  allow the flux level to vary through evenly spaced steps across the grid from (typically) 0.8--1.2 times the median observed flux density within each normalization band.  The exact range across the grid varies with each source and complex, to ensure that we sample sufficient parameter space surrounding the best-fit continuum level determined earlier.  We then systematically step through each continuum in the grid  and refit each complex for each source (following the same procedure as described earlier, but keeping the linear continuum fixed during each fit).  After each fit is performed, we calculate line properties  (e.g., $\lambda_c$, FWHM, line flux, \rew, etc.), and we record  $\chi^2_i$  (i.e., $\chi^2_i$ is for the fit using the $i^{\rm th}$ continuum model in the grid).  Since each fit within the grid contains the same number of degrees of freedom, we consider the \textit{relative} $\chi_i^2$ (i.e., $\Delta \chi^2_i = \chi^2_i - \chi^2_{\rm best}$) across the grid to estimate uncertainties.   The 68\% ($\sim$1$\sigma$) confidence interval typically corresponds to continua providing fits with $\Delta \chi^2_i \sim 30-50$ (given $\nu \sim$1900--6500, depending on the source and line complex).   For each continuum providing a $\Delta \chi^2_i$ within the 68\% confidence range, we determine the minimum and maximum value of each line property (i.e., $\lambda_c$, FWHM, \rew, etc.), linear continuum slope and intercept, and Fe continuum normalization  (when an iron template is included in the fit).  We use the minimum and maximum values of each parameter compared to the best-fit values derived earlier to determine approximate $\pm$1$\sigma$ uncertainties.  

\subsection{Notes on Fitting Individual Objects}
\label{sec:fits:notes}
{\it \srcone} ---  Narrow absorption systems for the \civ\ $\lambda \lambda$1548,1551 and \mgii\ $\lambda \lambda$ 2796,2804 doublets are detected at $z=1.733$.   When fitting the \ha\ complex, we mask 5800-6000~\AA\  to avoid potential broad \ion{He}{1} $\lambda$5877 emission blended with the continuum.

{\it \srctwo} --- Narrow absorption systems for \civ\ $\lambda \lambda$1548,1551 and \mgii\ $\lambda \lambda$ 2796,2804 are detected at $z=1.590$.  These systems were also reported by \citet{hryniewicz10} from the SDSS spectrum.     

{\it \srcthree} --- \oiii\ is detected in the NIR arm.  There is a hint of narrow \ha\ emission when visually examining this spectrum, but including a narrow \ha\ component does not improve the quality of the fit.

{\it \srcfive} --- We mask 5800-6000~\AA\ when fitting  \ha\ for this source.  A telluric line  falls near \mgii.  The affected pixels are given less weight during the spectral fitting, but this contamination may cause an additional source of systematic uncertainty on the measurements of the \mgii\ line.

{\it \srcsix} ---  We mask 5800-6000~\AA\ when fitting  \ha\ for this source.  A telluric line also falls near \mgii\ (see notes above).

{\it \srcseven} ---    \oiii\ is tentatively detected in the NIR arm.  There is a hint of narrow \ha\ emission when visually examining this spectrum, but including a narrow \ha\ component does not improve the quality of the fit.  There is  also a spurious feature in the UVB arm toward the blue end of the broad \civ\ emission line, near 3600~\AA\ in the observed frame.  This feature is an artifact from the reduction cascade, resulting from the continuum calibration lamp switching from a halogen lamp to a D$_2$ lamp at $\lambda \lesssim 3600$~\AA.   We therefore mask rest-frame 1485--1510~\AA\  prior to  fitting the \civ\ complex.   This artifact is not present in the other targets' UVB spectra. 

\tabletypesize{\footnotesize}
\tablecaption{Input Fitting Parameters}
\tablecolumns{6}
\tablewidth{0pt}
\renewcommand{\arraystretch}{1.2}
\begin{deluxetable*}{l l c c c c l}
\tablehead{
	\colhead{Name} & 
	\colhead{Fitting Window\tablenotemark{a}} & 
	\colhead{$\lambda_{\rm blue}$\tablenotemark{b}} & 
	\colhead{$\lambda_{\rm red}$\tablenotemark{b}} & 
	\colhead{Fe Template\tablenotemark{c}} & 
	\colhead{N$_{\rm broad}$\tablenotemark{d}} & 
	\colhead{Comments}  \\ 
	\colhead{} & 
	\colhead{(\AA)} & 
	\colhead{(\AA)} & 
	\colhead{(\AA)} & 
	\colhead{} & 
	\colhead{} & 
	\colhead{} \\ 
}
\startdata
\multicolumn{7}{l}{\ha\ Spectral Complex} \\ 
J0836  &  5600-6650\tablenotemark{e} & 5650  & 6200 & BG92 & 2 & \\
J0945  &  6000-6800 & 6200  & 6760 & BG92 & 3 &  \\
J1321  &  6000-7000 & 6200  & 6900 & \nodata & 2 & \\
J1411  &  5600-6600\tablenotemark{e} & 5650   & 6200 & \nodata & 1 &\\
J1417  &  5600-6700\tablenotemark{e}  & 5650  & 6200  & BG92 & 2 & \\
J1447  &  6000-7000 & 6200  & 6900 & \nodata & 2 & \\

\multicolumn{7}{l}{\hb\ Spectral Complex} \\ 
J0836 & 4425-4920 & 4435 & 4750 & BG92 & 2 & \\
J0945 & 4425-5040 & 4435 & 4750 & BG92 & 1 & \\
J1321 & 4425-5200 & 4435 & 5110 & BG92 & 2 & \oiii\ detected \\
J1411 & 4425-4910 & 4435 & 4750 & BG92 & 1 &  \\
J1417 & 4425-4980 & 4435 & 4750 & BG92 & 1 &  \\
J1447 & 4425-5200 & 4435 & 5110 & BG92 &  2 & \oiii\ tentatively detected \\

\multicolumn{7}{l}{\mgii\ Spectral Complex} \\ 
J0836 & 2600-3090 & 2655 & 3020 & VW01 & 1 & \mgii\ absorption doublet ($z=1.733$) \\
J0945 & 2200-3090 & 2250 & 3020 & VW01 & 1 & \mgii\ absorption doublet ($z=1.590$)\tablenotemark{f} \\
J1321 & 2600-3090 & 2655 & 3020 & VW01 & 1 & \\
J1411 & 2200-3090 & 2250 & 3020 & VW01 & 1 & \\
J1417 & 2200-3090 & 2250 & 3020 & VW01 & 1 & \\
J1447 & 2600-3090 & 2655 & 3020 & VW01 & 2 & \\

\multicolumn{7}{l}{\civ\ Spectral Complex} \\ 
J0836 & 1440-1700 & 1450 &  1690 & \nodata & 1 &  \civ\ absorption doublet ($z=1.733$)  \\
J0945 & 1440-1700 & 1450 &  1690 & \nodata & 1 & \civ\ absorption doublet ($z=1.590$)\tablenotemark{f} \\
J1321 & 1440-1700 & 1450 &  1690 & \nodata & 2 & \\
J1411 & 1440-1700 & 1450 &  1690 & \nodata & 1 & \\
J1417 & 1440-1700 & 1450 &  1690 & \nodata & 1 & \\
J1447\tablenotemark{g} & 1440-1700 & 1450 &  1690 & \nodata & 2 & \\

\enddata

\tablenotetext{a}{Rest-frame wavelength range for spectral fits.}
\tablenotetext{b}{Local linear continuum is normalized in two 20~\AA\ windows centered on $\lambda_{\rm blue}$ and $\lambda_{\rm red}$ (rest-frame).}
\tablenotetext{c}{Iron template used for each spectral complex.  BG92: \citet{boroson92}; VW01: \citet{vestergaard01}.}
\tablenotetext{d}{Number of broad Gaussian components in best-fit.}
\tablenotetext{e}{Rest-frame 5800-6000~\AA\ masked from fit, to avoid potential contamination from broad \ion{He}{1} $\lambda$5877.}
\tablenotetext{f}{Absorption doublet also detected in SDSS spectrum by \citet{hryniewicz10}.}
\tablenotetext{g}{Spurious feature from rest-frame 1485-1510~\AA\ is masked from the fit.  This feature is an artifact due to the calibration lamp switching from a halogen to a D$_2$  lamp at observed frame $\lambda \lesssim 3600$~\AA.}
\label{tab:infit}
\end{deluxetable*}
\renewcommand{\arraystretch}{1.}


\end{document}